\documentclass{aa}
\usepackage{graphicx}
\usepackage{natbib}
\def\cc{cm$^{-3}~$}
\def\ccc{cm$^{-3}$}

\def\brg{Br$\gamma$\/\ }

\def\av{A$_{\rm V}$\/\ }
\def\avv{A$_{\rm V}$}

\def\nee{{n$_e$}}
\def\hu{H75$\alpha$\/\ }
\def\hx{H92$\alpha$\/\ }

\def\l0{{l$_0$\/\ }}

\def\s{s$^{-1}$\/\ }
\def\sss{s$^{-1}$}
\def\nlyc{N$_{\rm Lyc}$\/\ }

\def\asec{$^{\prime\prime}$\/\ }
\def\asecc{$^{\prime\prime}$}
\def\amin{$^{\prime}$\/\ }
\def\aminn{$^{\prime}$}

\def\msunn{$M_{\odot}$}
\def\lsunn{$L_{\odot}$}
\def\ha{H$\alpha$\/\ }

\def\lsim{\raisebox{-0.3ex}{\mbox{$\stackrel{<}{_\sim} \,$}}}

\begin{document}
\title{Multi-density model of the ionised gas in NGC~253 using radio recombination lines}
\titlerunning{RRLs from NGC 253; multi-density \ion{H}{ii} gas}
\authorrunning{Mohan, Goss \& Anantharamaiah}
\author{Niruj R. Mohan \inst{1,2,3}
 \and W.M. Goss \inst{4}
 \and K.R. Anantharamaiah \thanks{Deceased.} \inst{2}}
\institute{
Institute d'Astrophysique de Paris, 98bis Boulevard Arago, Paris 75014, France \and
Raman Research Institute, C.V. Raman Avenue, Sadashivanagar Post Office, Bangalore 560080, India
\and Joint Astronomy Program, Department of Physics, Indian Institute of Science, Bangalore 560012, India
\and National Radio Astronomy Observatory, PO Box O, Socorro, NM 87801, USA
}
\offprints{W.M. Goss, \email{mgoss@aoc.nrao.edu}}
\date{Received  /  Accepted  }

\abstract{
We have imaged the \hx (8.3 GHz), \hu (15 GHz), and H166$\alpha$ (1.4 GHz) Radio Recombination 
Lines (RRLs) from NGC 253 at resolutions of 4.5 pc (0.4\asecc) , 2.5 pc (0.2\asecc) and 53 pc 
(4.5\asecc) respectively. The \hx line
arises from individual compact sources, most of which possess radio continuum counterparts. The line widths
range from $\sim$200 km \s for the sources near the radio nucleus to 70-100 km \s for the extranuclear
ones. These lines are emitted by gas at a density $\sim$10$^4$ \ccc. The remainder of the cm-wave RRLs
arise in lower density gas ($\sim$500 \ccc) with a higher area filling factor and with ten times higher
mass. A third component of higher density gas ($>$10$^4$ \ccc) is required to explain the mm-wave RRLs.
}

\maketitle

\section{Introduction} 

NGC 253 is a nearby (D=2.5 Mpc; 1\asecc=12 pc) highly inclined spiral galaxy,
undergoing a vigorous starburst (L$_{\rm FIR}$=3$\times$10$^{10}$ \lsunn; \citealt{th80})
in the central $\sim$100 pc region. The ongoing star formation is concentrated 
along an inclined ring. A number of compact near and mid-IR (NIR and MIR) sources 
have been discovered in this region
(Forbes, Ward, and Depoy 1991\nocite{forbes91}; Forbes et al. 1993\nocite{forbes93}; 
Pi\~na et al. 1992\nocite{pina92}; Keto et al. 1993\nocite{keto93}; Sams et al. 
1994\nocite{sams94}). Optical imaging by Watson et al. (1996)\nocite{watson96}
revealed the presence of four star clusters, which were identified with individual MIR/NIR sources
(see also Forbes et al. 2000\nocite{forbes00}). The morphology of the radio continuum 
emission, which is unaffected by dust extinction, is quite different. Turner and Ho 
(1985)\nocite{th85} discovered a linear chain of compact sources at 15 GHz, which were studied 
in detail by Antonucci and Ulvestad (1988)\nocite{au88} and Ulvestad and Antonucci 
(1997; hereafter UA97)\nocite{ua97}. The latter were able to classify the brighter 
compact sources as synchrotron or free-free dominated, based on the derived spectral indices.
Kalas and Wynn-Williams (1994)\nocite{kww94} and Sams et al (1994)\nocite{sams94} showed that 
most of the infrared sources are regions of low dust extinction and are not coincident
with the radio sources. Hence, the radio sources and the infrared-optical sources seem to trace
very different types of objects.
In particular, the peak of the NIR emission is offset by about 3.5\asec southwest from the
peak of the radio continuum emission (the radio nucleus). There is no
associated radio emission towards the NIR peak and only weak NIR emission is observed
near the radio nucleus. High resolution radio continuum observations with the Very Large Array (VLA)
suggest that the radio nucleus hosts an Active Galactic Nucleus (AGN; \citealt{th85}, UA97\nocite{ua97}), 
which was subsequently confirmed (\citealt{mohan02}, \citealt{weaver02}). The NIR continuum peak, identified 
as a region of active star formation, also coincides with the position of the 
emission peaks of various ionised gas tracers ([NeII]: B\"oker, Krabbe
and Storey 1998\nocite{boker98}; Keto et al. 1999\nocite{keto99}; \brg: Forbes et al.
1993\nocite{forbes93}; optical continuum and line
emission: Engelbracht et al. 1998\nocite{engel98}; Watson et al. 1996\nocite{watson96};
Forbes et al. 2000\nocite{forbes00}).

The immediate product of the starburst process, with its associated high star formation rate 
and efficiency, is a young massive stellar population. This population can studied through its
surrounding ionised gas, the properties of which are determined partly by the ionising stars themselves,
and partly by gas dynamics. Recent star formation is well traced by NIR emission, except in regions of 
high dust extinction. However, such highly obscured star formation can be probed by radio continuum 
and hydrogen radio recombination line (RRL) emission (except in cases where the continuum optical depth 
due to free-free emission is much larger than unity). RRL emission is insensitive to dust extinction, arises 
only in thermal gas, and can be imaged at arcsecond resolution. 
NGC 253 was the first extragalactic object (excluding the Magellanic Clouds) to 
have been detected in RRLs \citep{sb77}. This line emission was initially modelled as being stimulated
by the background nuclear emission \citep{shaver78,mebold80}. 
The first interferometric observations of RRLs from NGC 253 were carried out by \citet{ag90} at 
higher frequencies (the H166$\alpha$, H110$\alpha$ and the H92$\alpha$ lines at 1.4 GHz,
4.8 GHz and 8.3 GHz respectively) using the VLA. Models indicated that these lines arise due
to internal emission from high density gas (\nee$\sim$10$^4$ \ccc; similar to compact 
HII regions). \citet{puxley97} observed mm-wave RRLs from this galaxy and suggested that the origin of
the RRL emission at all frequencies is spontaneous emission from optically 
thin HII regions. A total ionisation rate of (3.7 $\pm$ 0.8)$\times$10$^{53}$~\s 
was derived, an order of magnitude higher than that 
derived from infrared emission lines. The ionisation budget and hence the star formation rate as
determined by RRLs depends on the nature of the line emission, which is not yet fully understood.
High resolution observations ($\sim$1.3\asec or 16 pc) 
of the 8.3 GHz RRL from NGC 253 were carried out by \citet{ag96}. They showed that the line emission 
peaks at the radio nucleus (as do the higher frequency RRLs as well; \citealt{zhao01}). They 
detected much weaker line emission near the NIR peak, which is unexpected, since
the infrared peak is probably the most active star forming site in this galaxy.
Using the high resolution \hx and \hu data presented in this work, the RRL emission from the radio 
nucleus alone was studied separately and the results were presented by \citealt{mohan02}.
Based on the results of the line emission 
models, it was concluded that the ionising source for the RRL emitting gas is probably
a central low luminosity AGN with weak X-ray emission, which was subsequently confirmed through
$CHANDRA$ observations \citep{weaver02}.

In order to investigate the nature of RRL emission in this galaxy, and hence the nature of
obscured star formation, we have performed high resolution observations of RRLs in
NGC 253 at multiple frequencies using the VLA. The 8.3 GHz H92$\alpha$ line and the 15 GHz H75$\alpha$ 
line were observed using the VLA in the A configuration, which constitute the highest resolution 
RRL observations to date (0.4\asec and 0.2\asec respectively). These line emissions from the
radio nucleus were studied separately and have been described in detail in \citet{mohan02}. The current
paper deals with the remainder of this cm-wave RRL emission. We have carried out these observations 
in order to (1) locate the sites of high frequency RRL emission and identify radio continuum 
counterparts to the line emitting sources, (2) to investigate the
nature of the radio nucleus and the IR peak, based on their RRL emission, and (3) separate the
diffuse and compact RRL emission in an attempt to construct a multi-density model of the ionised
gas in NGC 253, using multi-frequency RRL data, similar to Arp 220 \citep{anan00}. The 1.4 GHz 
H166$\alpha$ line emission was observed at a resolution of $\sim$4\asec using the VLA in the 
BnA configuration and the C configuration data presented in \citet{ag90} were re-analysed. 
The aim was to better localise the large-scale line emission and hence (1) constrain the 
projected lateral size of the line emitting gas and thus obtain more accurate models, 
(2) investigate the relative importance of externally stimulated emission versus 
internal (spontaneous) emission at 1.4 GHz compared to higher frequencies and (3) check the 
reality of the observed position offset between the low frequency and high frequency
RRL emission found by \citet{ag90} using the C and D configuration data. 

\section{Observations}

\begin{table*}
\begin{center}	
\caption[]{\bf VLA observational log and image parameters}
\label{tobspara}
\small
\begin{tabular}{lccc}
\hline
\multicolumn{1}{l}{Parameter}&\multicolumn{1}{c}{1.4 GHz}&\multicolumn{1}{c}{8.3 GHz}&\multicolumn{1}{c}{15 GHz} \\
\hline
\multicolumn{1}{l}{VLA configuration} & \multicolumn{1}{c}{BnA} & \multicolumn{1}{c}{A} & \multicolumn{1}{c}{A} \\
Date of observation             &  01 Oct 95 & 9 \& 12 Jul, & 26 Jun 99 \\
                                &           &  9 Oct 99  & \\
$\nu_{\rm rest}$ of RRL (MHz)   &  1424.7(H166$\alpha$) & 8309.4  &  15281.5 \\
                                &  1374.6(H168$\alpha$) & (H92$\alpha$) & (H75$\alpha$) \\
Bandwidth (MHz)                 &  3.03 & 24.2 & 46.9 \\
\# of channels, \# IF           &  31, 4 & 31, 1 & 15, 1\\
Spectral resolution$^a$\,\,(km \sss)   & 41.1 & 56.4 & 122.6 \\
Shortest baseline (k$\lambda$)  & 2.4 & 7 & 12 \\
Synthesized beam (\asecc$\times$\asecc) & 5.07$\times$3.79 & 0.50$\times$0.28 & 0.31$\times$0.14 \\
Phase calibrator                &  0023-263 & 0116-219 & 0118-272 \\
Bandpass calibrator             &  0023-263,3C48 & 2251+158 & 2251+158 \\
\hline
Peak continuum\,\,\,(mJy/beam) &  675 & 40 & 37  \\
Continuum noise (mJy/beam) & 0.13 & 0.04 & 0.13 \\
Noise per channel (mJy/beam)      &  0.26 & 0.12 & 0.35 \\
\hline
\multicolumn{4}{p{3in}}{$^a$\scriptsize{The spectral resolution after off-line Hanning smoothing.}} \\
\end{tabular}
\end{center}
\end{table*}

\subsection{Centimetre-wave RRLs: 8.3 GHz and 15 GHz observations} 

The 8.3 GHz \hx and the 15 GHz \hu recombination lines from NGC 253 were observed
using the VLA in the A configuration and the observational parameters are listed in 
Table~\ref{tobspara}. Only part of the 8.3 GHz dataset was used to study the weak 
line emission since the remaining data were affected by the 3 MHz ripple in the VLA waveguide
system\footnote{See VLA Test Memorandum 158 by C.L. Carilli.} (however, \citealt{mohan02}
used all of the data since the much stronger nuclear line emission is far less affected
by the ripple). All data analysis were made using standard algorithms available within the
software AIPS. The continuum emission at 8.3 GHz and 15 GHz was imaged. Offline Hanning smoothing 
was applied to the line data to reduce the effects of Gibb's ringing \citep{bible}. 
The line emission at both
frequencies were imaged after applying corrections for the observed positional offsets 
(see Sect.~\ref{sec_posn}). 

\subsection{Decimetre-wave RRLs: 1.4 GHz data} 

NGC 253 was observed at 1.4 GHz using the VLA in the BnA configuration
and the observational details are listed in Table~\ref{tobspara}. The archival C configuration data
were reprocessed and are described in the appendix.
The BnA array observations were performed in the 4 IF mode of the correlator in order to observe
both the H166$\alpha$ and the H168$\alpha$ lines simultaneously. The continuum emission was imaged 
and deconvolved simultaneously with emission from strong sources within the primary beam of 30\aminn. 
The continuum emission from NGC 253 is surrounded by negative bowls of emission due to missing short spacing 
visibilities. Bandpass calibration every half hour proved adequate to track the time variation of the 
bandshape to an accuracy of $<$1 \%. Both the H166$\alpha$ and the H168$\alpha$ lines were detected in the 
line cubes and the line parameters were found to be consistent in the two datasets.

\section{Results}

\subsection{Centimetre-wave RRLs: 8.3 GHz and 15 GHz observations} 
\label{sec_posn}

\begin{figure}[h]
\resizebox{\hsize}{!}{\includegraphics{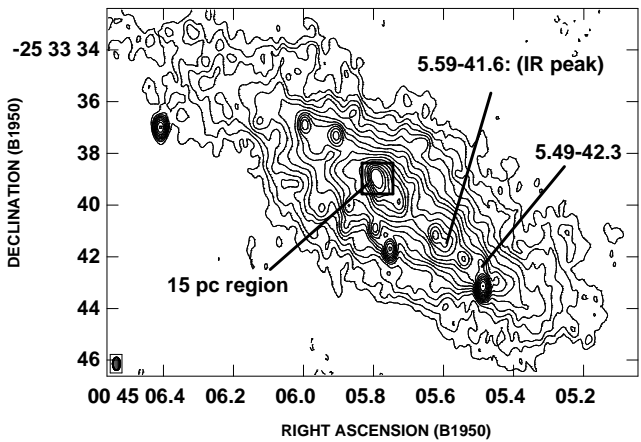}}
 \caption{8.3 GHz continuum emission from the central disk of NGC 253 using VLA A
configuration data. The rms in the image is
0.04 mJy/beam and the peak flux density is 63 mJy/beam. The contour levels are (-5, 5, and
higher in steps of 1.4) times the rms. The synthesized beam is
0.50\asecc$\times$0.28\asecc, P.A.=0$^{\circ}$ (6 pc $\times$3.4 pc). 
The sources 5.49$-$42.3, 5.59$-$41.6 and the 15 pc region marked in the 
figure refer to \hx line emitting regions, discussed in Sect.~\ref{sec_morphX}.
The two remaining line sources 5.72$-$40.1 and 5.75$-$39.9 cannot be seen clearly in 
the figure.  }
\label{fXcont}
\end{figure}

\begin{figure}[h]
\resizebox{\hsize}{!}{\includegraphics{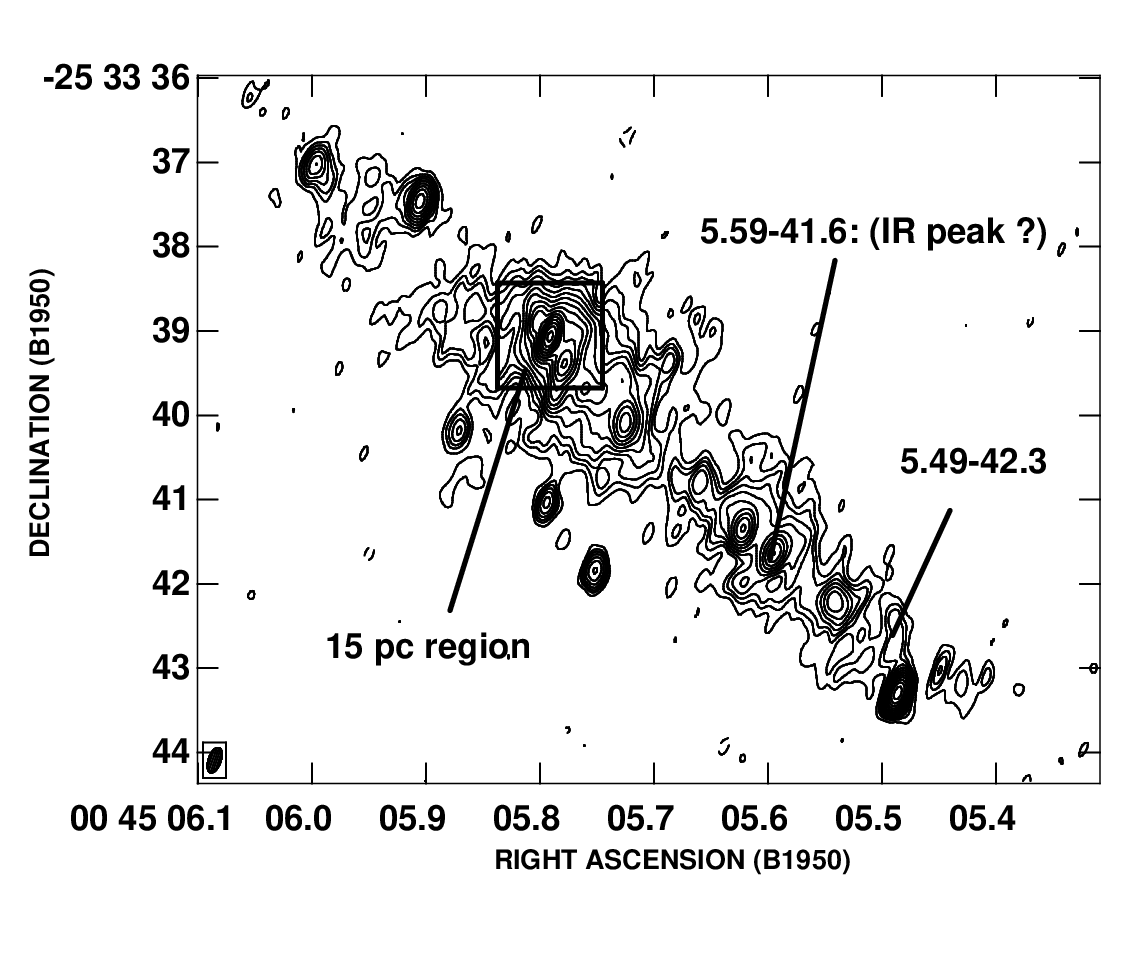}}
 \caption{15 GHz continuum emission from the central disk of NGC 253 using VLA A
configuration data. The rms in the image is
0.12 mJy/beam and the peak flux density is 44 mJy/beam. The contour levels are (-3, 3, 5, and 
higher in steps of 1.35) times the rms.
The synthesized beam is 0.31\asecc$\times$0.14\asecc, P.A.=$-$20$^{\circ}$~(3.7~pc $\times$1.8 pc).
The sources 5.49$-$42.3, 5.59$-$41.6 and the 15 pc region marked in the 
figure refer to \hx line emitting regions, discussed in Sect.~\ref{sec_morphX}.
The two remaining line sources 5.72$-$40.1 and 5.75$-$39.9 cannot be seen clearly in 
the figure.  }
\label{fUcont}
\end{figure}

The continuum images made at both 8.3 GHz and 15 GHz are consistent with the 
images published by UA97\nocite{ua97}; the continuum flux densities of individual compact 
sources agree to within 10 \%. However, the positions of these sources in
the 8.3 GHz and 15 GHz continuum images do not coincide. This positional discrepancy 
can be modelled as a relative global offset, indicating the uncertainty in the positions of the phase 
calibrators as a probable cause. Adopting the positions of compact sources in the A 
configuration 15 GHz image of UA97\nocite{ua97}, the positional offsets were derived to be 
+0.036\asecc$\pm$0.01\asec in R.A. and +0.14\asecc$\pm$0.01\asec in Dec for the 8.3 GHz data, and 
+0.021\asecc$\pm$0.007\asec in R.A. and +0.012\asecc$\pm$0.006\asec in Dec for 
the 15 GHz data, which were then applied to the data. All positions quoted in this paper are 
relative to 5.75$-$41.8\footnote{The corrected (B1950.0) coordinates of 5.75$-$41.8 are 
00$^{\rm h}$45$^{\rm m}$05$^{\rm s}$.751, $-$25$^{\circ}$33\aminn41$^{\prime\prime}$.85 
at 8.3 GHz and 00$^{\rm h}$45$^{\rm m}$05$^{\rm s}$.750; $-$25$^{\circ}$33\aminn41$^{\prime\prime}$.86
at 15 GHz.}. The images of continuum emission at the two frequencies are shown in Fig.~\ref{fXcont} and
Fig.~\ref{fUcont}. 

\begin{figure}
 \includegraphics[width=3.4in]{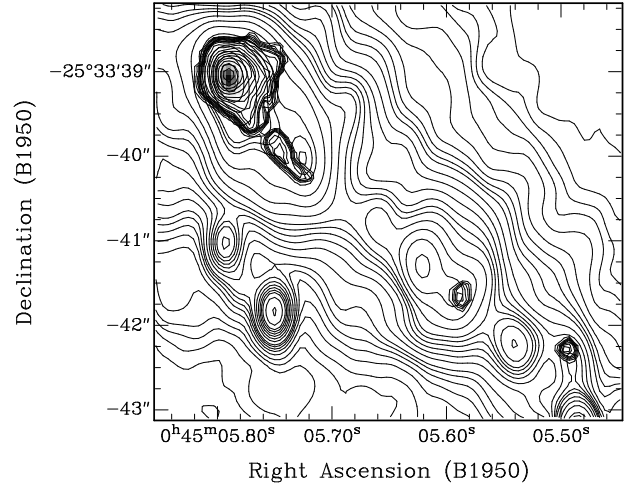} \\
 \includegraphics[width=3.4in]{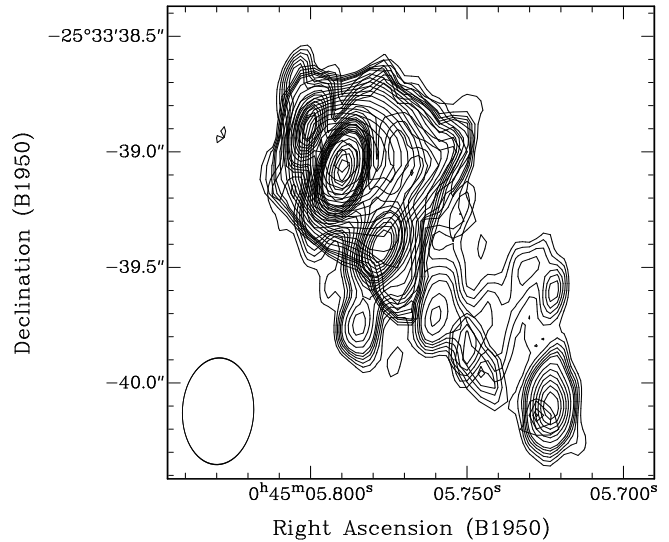}
 \caption{The four extranuclear \hx sources, discussed in Sect.~\ref{sec_morphxnuc}, and the central 
extended emission are detected. {\bf Top:} The \hx line emission (thick contours) overlayed on the 
8.3~GHz continuum emission (thin contours). Two compact sources of 
line emission (5.49$-$42.3 and 5.59$-$41.6) are seen towards the SW of the emission surrounding 
the radio nucleus. The synthesized beam is 0.50\asecc$\times$0.28\asecc, P.A.=0$^{\circ}$ and rms of the continuum
image is 0.04 mJy/beam. 
{\bf Bottom:} A close-up view of the \hx line emission towards the 25 pc region (grey scale)
overlayed on a high resolution 8.3~GHz continuum image (contours). The line emission consists of two
additional extranuclear sources (5.72$-$40.1 and 5.75$-$39.9) and a contiguous 15 pc sized 
central line emitting region. The synthesized beam of the continuum and line images are 0.35\asecc$\times$0.19\asecc,
P.A.=$-$3$^{\circ}$ and 0.45\asecc$\times$0.24\asecc, P.A.=$-$3$^{\circ}$ respectively. The rms of the continuum
image is 0.07 mJy/beam and the contours are in steps of 1.4 starting from 10$\sigma$. }
\label{fXallmom}
\end{figure}

The integrated \hx emission image using the full 
resolution dataset is shown in Fig.~\ref{fXallmom}a overlaid on an image of the continuum 
emission from the region. The emission from the inner 25 pc region is shown in
Fig.~\ref{fXallmom}b, overlaid on a higher resolution 15 GHz continuum image. 
The line emission is spatially extended, with the peak of the emission coinciding with the peak 
of the radio continuum (radio nucleus). This nuclear line emission was discussed in detail in 
\citet{mohan02}. The strongest line emission arises in an extended region surrounding the nucleus,
spread over 15 pc. In addition, multiple compact line emitting sources were found to lie along 
the larger continuum disk. 
The morphology of the
\hx line emission is quite complex and is discussed in detail in Sect.~\ref{sec_morphX}.
The velocity field of the ionised gas in the 15~pc region indicates that the velocity gradient is almost in 
the east-west direction and the value increased from 200 to 230 km \s from east to west, consistent with 
the observations of \citet{ag96}. The analysis of the velocity field is not undertaken in this paper.

\begin{figure}
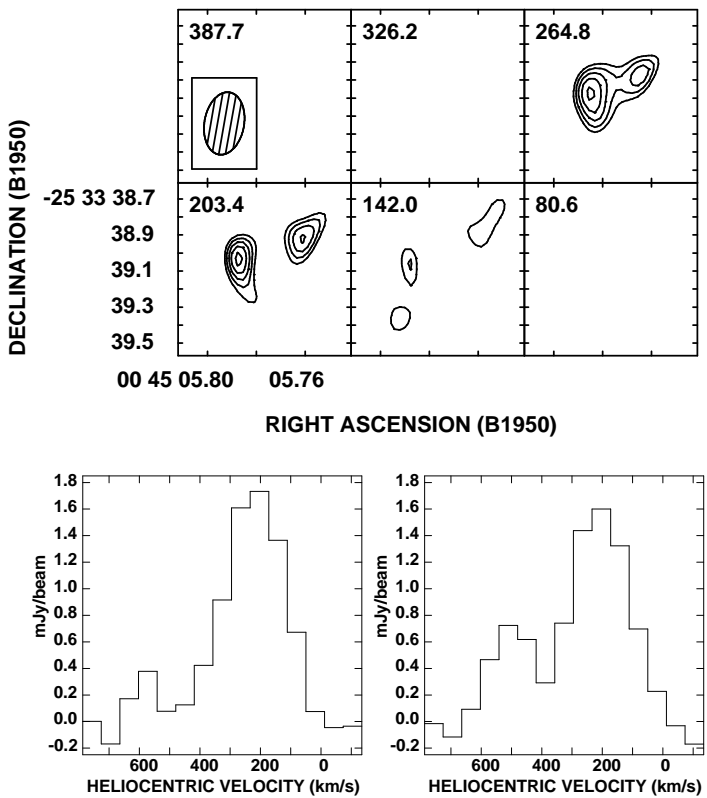

\resizebox{\hsize}{!}{\includegraphics[width=1in]{fig253_5.ps}} \\
\resizebox{\hsize}{!}{\includegraphics{fig253_6.ps} $~~~~~~~~~~~$\includegraphics{fig253_7.ps}}
 \caption{{\bf (a).} VLA A configuration `naturally weighted' images of \hu emission in individual
channels from the central 15 pc region of NGC 253. The central
heliocentric velocity (in km s$^{-1}$) of each channel is marked in each frame. 
The synthesized beam is 0.35\asecc$\times$0.22\asecc, P.A.=$-$10$^{\circ}$ and the rms
in the channel images is 0.36 mJy/beam. The contour levels are (3.5, 3.8, 4.1, 4.4, 4.7,
5, 5.3) times the rms.
{\bf (b).} \hu spectrum towards the continuum peak (radio nucleus).
{\bf (c).} \hu spectrum towards the line source 5.76-38.9. 
}
\label{chmapU}
\end{figure}

The 15 GHz line cube was used to search for sources of H75$\alpha$ line 
emission. The channel images of the \hu line emission as well the 
\hu spectra towards the two sources are shown in Fig.~\ref{chmapU}.
The peak of the line emission is coincident with the radio nucleus and
the only other source of RRL emission arises from a compact source 
$\sim$0.35\asec northwest of the nucleus. This source seems to have corresponding 
\hx line emission at 8.3 GHz, associated with the continuum source 5.76$-$38.9 
(Sect.~\ref{sec_morphX}), and is the second strongest line emitting source at 
the latter frequency. 

\subsection{Decimetre-wave RRLs: 1.4 GHz data} 

\begin{figure}
\resizebox{\hsize}{!}{\includegraphics[angle=-90]{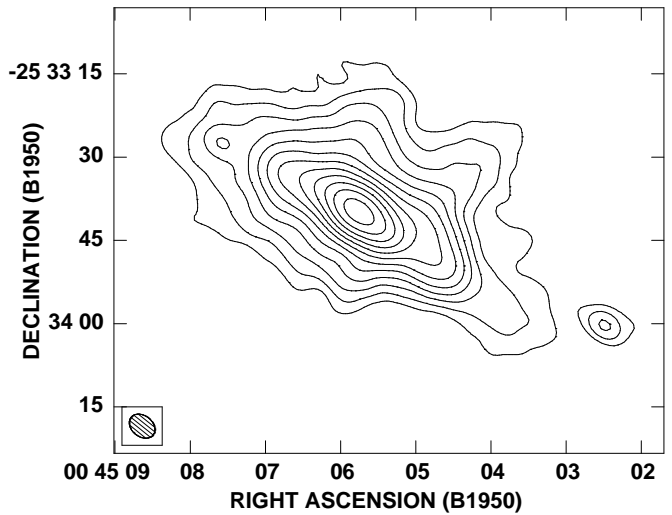}}
 \caption{1.4~GHz continuum emission from the inner disk of NGC 253 using the 
VLA BnA configuration data. The rms in the image is 0.13~mJy/beam and the contour levels are from 
1.3 to 664~mJy/beam in steps of 1.7. The synthesized beam is 5.1\asecc$\times$3.8\asecc, 
P.A.=51$^{\circ}$.}
\label{fLcontB}
\end{figure}

\begin{figure}
\resizebox{\hsize}{!}{\includegraphics{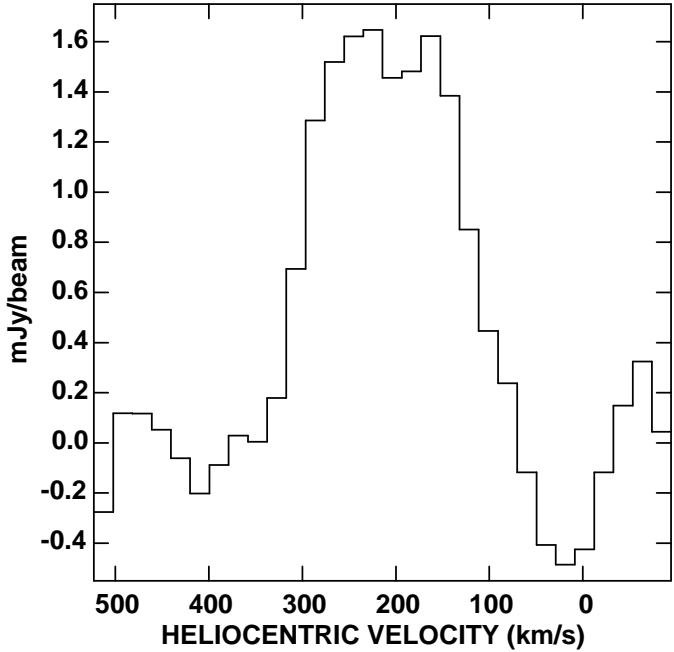}}
 \caption{Co-added H166$\alpha$ and H168$\alpha$ spectra towards the continuum peak of NGC 253 using
the VLA BnA configuration data for the natural weighted image. Synthesized beam is 
5.1\asecc$\times$3.8\asecc, P.A.=50$^{\circ}$ and the rms noise in the line cube is 0.26 mJy/beam.}
\label{fLlineB}
\end{figure}

The 1.4 GHz continuum image is shown in Fig.~\ref{fLcontB}. 
The data for both the RRLs were combined and the line emission was imaged after 
offline Hanning smoothing. The line emission coincides with the position of the 
peak of the continuum emission and is marginally spatially resolved. The line 
does not increase in strength by more than 10 \% in images made at lower resolution. Hence the 
synthesized beam (5.1\asecc$\times$3.8\asecc) covers almost all of the line emitting region.
The spectrum obtained from the combined dataset towards the continuum peak
is shown in Fig.~\ref{fLlineB}. The line parameters are listed in Table~\ref{talllinepara}.

\begin{table*}
\begin{center} 
\caption[]{\bf H168$\alpha$, \hx and \hu line parameters for NGC 253}
\label{talllinepara}
\begin{tabular}{lcccc}
\hline
\multicolumn{1}{l}{Source} & \multicolumn{1}{c}{Peak flux} & \multicolumn{1}{c}{Central velocity$^a$} & 
\multicolumn{1}{c}{FWHM$^b$} & \multicolumn{1}{c}{Integrated flux} \\
\multicolumn{1}{l}{ } & \multicolumn{1}{c}{density (mJy/beam)} & \multicolumn{1}{c}{(km \sss)} & 
\multicolumn{1}{c}{(km \sss)} & \multicolumn{1}{c}{($\times$10$^{-23}$ W m$^{-2}$)} \\
\hline
\multicolumn{5}{c}{\em 1.4 GHz H168$\alpha$ line emission (BnA configuration)}\\
\hline
7.4\asecc$\times$5.2\asec area & 1.8 $\pm$ 0.1 & 200 $\pm$ 20 & 160 $\pm$ 14 & 1.6 $\pm$ 0.25 \\
\hline
\multicolumn{5}{c}{\em 8.3 GHz \hx line emission$^c$ (A configuration)}\\
\hline
5.49$-$42.3             & 0.5 $\pm$ 0.1   & 201 $\pm$ 7  & 73 $\pm$ 17  & 1.2 $\pm$ 0.3 \\
5.59$-$41.6             & 0.5 $\pm$ 0.1   & 172 $\pm$ 7  & 65 $\pm$ 16  & 1.0 $\pm$ 0.3 \\
5.72$-$40.1             & 0.46 $\pm$ 0.09 & 170 $\pm$ 15 & 101 $\pm$ 23 & 1.4 $\pm$ 0.3 \\
5.75$-$39.9             & 0.54 $\pm$ 0.09 & 175 $\pm$ 10 & 112 $\pm$ 20 & 1.8 $\pm$ 0.3 \\
`15~pc region'          & 3.2  $\pm$ 0.15 & 194 $\pm$ 5  & 211 $\pm$ 12 & 20 $\pm$ 2    \\
\hline
\multicolumn{5}{c}{\em 15 GHz \hu line emission$^c$ (A configuration)}\\
\hline
Nucleus  & 1.8 $\pm$ 0.3 & 224 $\pm$ 17 & 230 $\pm$ 40 & 23 $\pm$ 3 \\
5.76$-$38.9 & 1.6 $\pm$ 0.3 & 205 $\pm$ 18 & 230 $\pm$ 43 & 20 $\pm$ 3 \\
\hline
\multicolumn{5}{p{3in}}{$^a$ Heliocentric velocity, optical definition.}\\
\multicolumn{5}{p{5in}}{$^b$ Corrections for finite velocity resolution not applied.}\\
\multicolumn{5}{p{5in}}{$^c$ All sources identified by coordinates are spatially unresolved. 
See Table~\ref{tobspara} for synthesized beam sizes. }\\ 
\end{tabular}
\end{center}
\end{table*}

The RRL emission from NGC 253 was imaged by \citet{ag90} at low resolution, using
the VLA in the C and D configurations. The 1.4 GHz line emission was found to be
shifted SE from the peak of the continuum and the 4.8 GHz and 8.3 GHz line emissions. 
Subsequently, the D configuration data were found to exhibit problems in the spectral domain and
was hence unusable. 
The reanalysed C configuration data (appendix) shows the line
emission to be centred on the continuum peak. Hence the positional offset for the 
1.4 GHz line emission presented earlier by  \citet{ag90} is not confirmed.

\section{The morphology and analysis of the extended \hx RRL emission}
\label{sec_morphX}

In this section, the morphology of the extended \hx line emission is analysed in order to 
identify the individual line emitting sources and derive their properties.
Continuum counterparts can be associated with most, but not all, of the line emitting 
sources. We identify each of the RRL sources by the continuum counterpart or the nearest 
continuum source. There are two compact line sources to the SW (Fig.~\ref{fXallmom}a), 
and the central region consists of two additional compact sources along with the extended 
15 pc sized nuclear emission (Fig.~\ref{fXallmom}b). These four extranuclear line sources
are described in the next section, followed by a discussion of the emission from the central
15 pc region.

\subsection{The extranuclear \hx sources}
\label{sec_morphxnuc}

\begin{figure}[h]
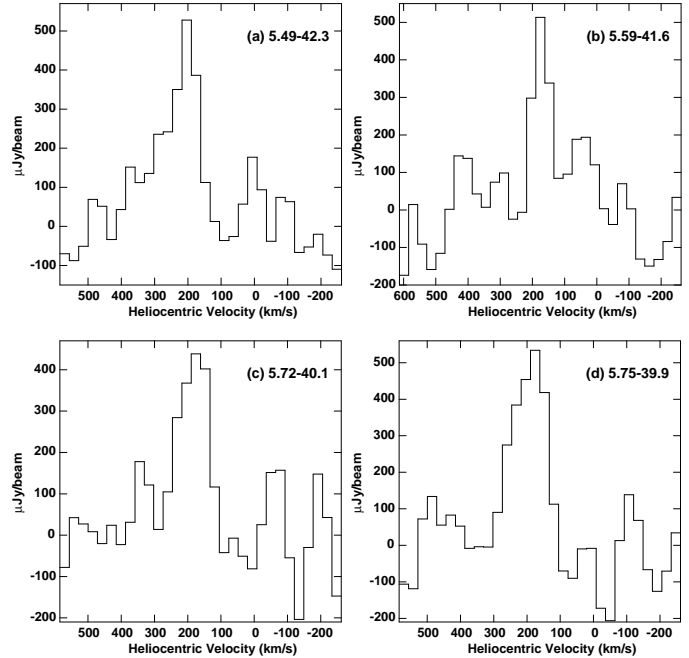

\resizebox{\hsize}{!}{\includegraphics{fig253_10.ps} $~~~~~~~~$\includegraphics{fig253_11.ps}}
\resizebox{\hsize}{!}{\includegraphics{fig253_12.ps} $~~~~~~~~$\includegraphics{fig253_13.ps}}
 \caption{\hx VLA spectrum towards the four compact extranuclear line sources using the A 
configuration data, identified by the coordinates of the associated/nearest continuum source :
{\bf (a).} 5.49$-$42.3, {\bf (b).} 5.59$-$41.6, {\bf (c).} 5.72$-$40.1 and {\bf (d).} 5.75$-$39.9.
The synthesized beam is 0.50\asecc$\times$0.28\asecc, P.A.=$-$2$^{\circ}$. The rms in the line cubes is
0.12 mJy/beam and the velocity resolution is 56.4 km \s.}
\label{specxnuc}
\end{figure}

The western most line source (seen in Fig.~\ref{fXallmom}a) is nearly coincident with the 
continuum source 5.49$-$42.3. A
significant fraction of its continuum emission is thermal (the 8.3 GHz--23 GHz spectral index 
is $-$0.28 $\pm$ 0.36; UA97\nocite{ua97}), supporting its association with the line emission.
The line emitting region 1.5\asec east of this source (Fig.~\ref{fXallmom}a) is displaced 
$\sim$0.1\asec southwest of the compact continuum source 5.59$-$41.6. 
This object is the second most prominent RRL source in the lower resolution image 
of \citet{ag96} and is probably associated with the IR peak (see Sect.s~\ref{sec_hydenmodel} 
and \ref{sec_irpeak}). Two additional sources of compact \hx line emission are detected 
in the central region (Fig.~\ref{fXallmom}b).
The western source is close to the continuum source 5.72$-$40.1. The latter exhibits a 
flat spectral index from 4.8 GHz up to 23 GHz and was identified by UA97\nocite{ua97} 
as the brightest thermal source in NGC 253.  The eastern source, however, has no obvious 
continuum counterpart, and in fact lies in a region of minimum continuum emission. 
Nevertheless, the nearest continuum source 5.75$-$39.9 will be used to refer to this 
line emitting source. The \hx 
spectra towards these extranuclear line sources are presented in Fig.~\ref{specxnuc}. 
These line sources are spatially unresolved and the spectral parameters are listed in 
Table~\ref{talllinepara}.

\subsection{The \hx emission in the central 15 pc region}

\begin{figure*}
 \includegraphics[width=5.5in]{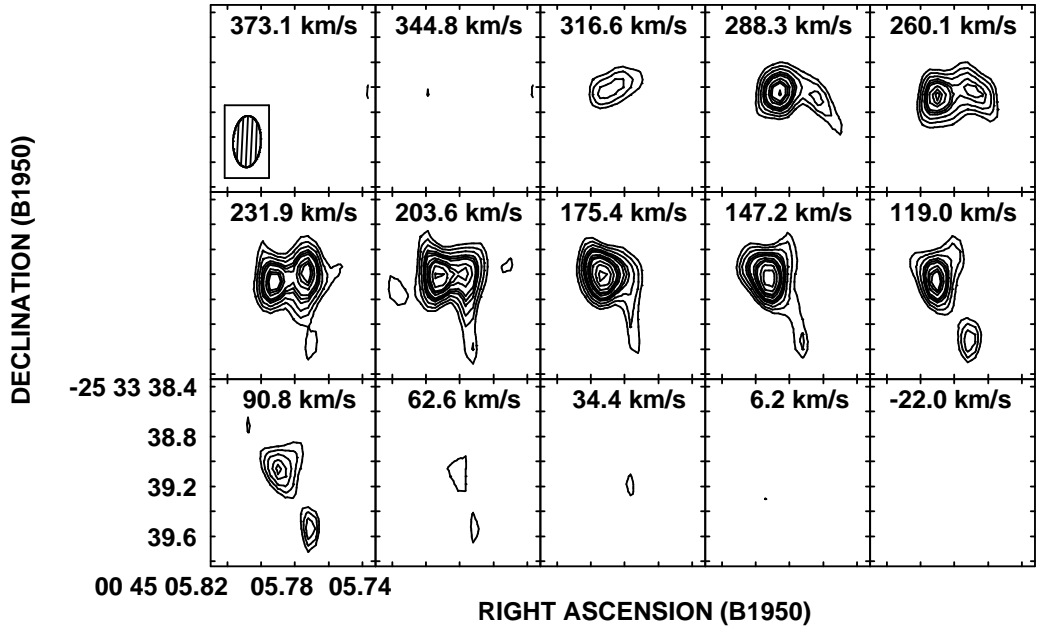}
 \caption{VLA A configuration images of the \hx line emission in individual spectral channels 
in the central `15 pc region' of NGC 253, at higher resolution. The central heliocentric 
velocity of each channel is marked in each frame. Line emission in every alternate 
channel is independent due to Hanning smoothing.
The synthesized beam is 0.41\asecc$\times$0.22\asecc, P.A.=10$^{\circ}$ and 
the rms in the channel images is 0.13 mJy/beam. The contour levels are 
(-3, 3, 3.5, 4, 4.5, 4.75, 5, 5.5, 5.8, 6, 6.5, 7, 7.5) times the rms.
}
\label{chmapX}
\end{figure*}

\citet{ag96} found that the strongest \hx emission arises from the region surrounding the 
nucleus. Our observations, which have three times improved angular resolution, 
can spatially resolve this central emission, which peaks at the radio nucleus. High 
resolution channel images of this 15~pc region 
are shown in Fig.~\ref{chmapX}. The line emission in this region can be resolved into four 
discrete components, as can be seen in Fig.~\ref{chmapX}, 
though the identification of the individual sources is uncertain due to the complex 
morphology and the low signal to noise. However, each of the sources can be best identified 
in different channel images. 
One or more channels were identified where the sources can be distinguished from the 
surrounding emission (the southern source at a velocity of 119.0 and 90.8 km \sss, 
the westernmost at 203.6 km \s and the two northern sources at 231.9 and 260.1 km \sss).
These channels were used to derive the peak positions by fitting spatially unresolved gaussians
using the AIPS task JMFIT. The errors in the positions were determined by the fitting procedure.
Given the low signal-to-noise of the emission (the peak emission is in the range of 5-9~$\sigma$), 
a simultaneous 3-d deconvolution of all four line sources was not attempted. 

\begin{table*}
\begin{center}
\caption[]{\bf Positions of the \hx and \hu line emitting sources in the 15 pc region}
\label{tlineXnucposn}
\begin{tabular}{ccc}
\hline
\multicolumn{1}{c}{Source} & \multicolumn{1}{c}{RA} & \multicolumn{1}{c}{Dec} \\
\multicolumn{1}{c}{ } & \multicolumn{1}{c}{(B1950.0)} & \multicolumn{1}{c}{(B1950.0)} \\
\hline
\multicolumn{3}{c}{\em The four \hx sources (see Fig.~\ref{mom0nuc})}\\
\hline
1  & 0$^{\rm s}$.002 (2) & 2\asecc.90 (5) \\
2  & 0$^{\rm s}$.022 (2) & 2\asecc.31 (5)\\
3  & 0$^{\rm s}$.026 (1) & 2\asecc.87 (2) \\
4  & 0$^{\rm s}$.044 (1) & 2\asecc.80 (2) \\
\hline
\multicolumn{3}{c}{\em The second \hu source}\\
\hline
5.76$-$38.9 & 0$^{\rm s}$.019 (2) & -2\asecc.94 (5)
\\
\hline
\multicolumn{3}{p{3in}}{\scriptsize NOTE: The coordinates are with respect to 
00$^{\rm h}$45$^{\rm m}$05$^{\rm s}$.751; $-$25$^{\circ}$33\aminn41$^{\prime\prime}$.85,
which is the (B1950.0) position of 5.75$-$41.8 in the 8.3 GHz image after applying 
positional corrections (Sect.~\ref{sec_posn}). Positive offsets are for eastward
and northward directions.
The gaussian fitting errors in the final decimal place are given in parenthesis.}\\
\end{tabular}
\end{center}
\end{table*}

\begin{figure}
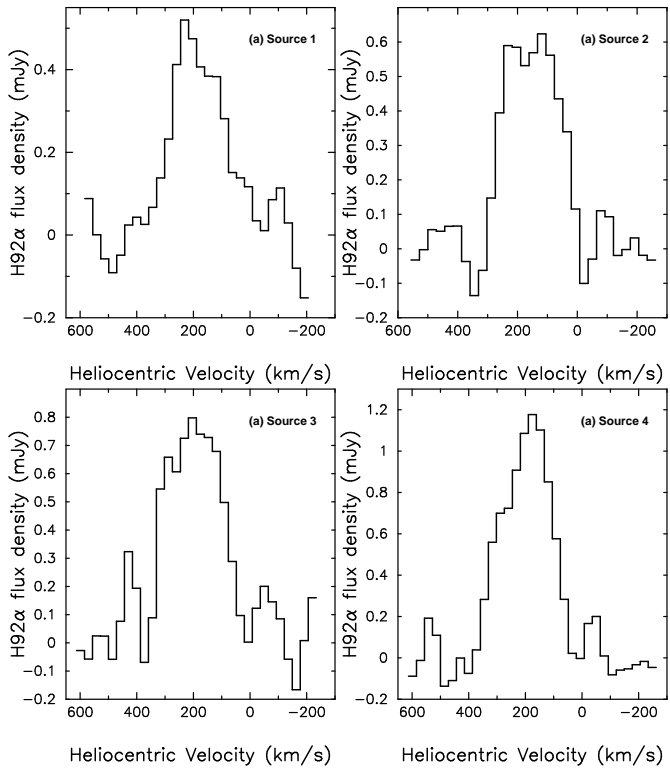

\resizebox{\hsize}{!}{\includegraphics{fig253_17.ps} \includegraphics{fig253_18.ps}}
\resizebox{\hsize}{!}{\includegraphics{fig253_19.ps} \includegraphics{fig253_20.ps}}
 \caption{The VLA A configuration \hx spectra towards the line emitting sources in the 15~pc region, obtained towards
the positions listed in Table~\ref{tlineXnucposn}. The synthesized beam is 0.51\asecc$\times$0.28\asecc, 
P.A.=$-$2$^{\circ}$. The rms noise in the channel images is 0.12 mJy/beam.}
\label{specXnuc}
\end{figure}

The derived positions of these four sources are listed in Table~\ref{tlineXnucposn} 
and are also marked in Fig.~\ref{mom0nuc}a, which shows a high resolution image of the integrated line emission 
from this region. The offsets between the marked positions and those of the line peaks in the figure are 
due to blending of emission from the four sources. In Fig.~\ref{mom0nuc}b, the integrated line emission is overlaid 
on a high resolution 15 GHz continuum image. Apart from the nucleus, the source
immediately to the west coincides with a strong continuum source that is not listed in the
catalogues of UA97\nocite{ua97}. This \hx line source also has a \hu line emission counterpart 
(Fig.~\ref{chmapU}). The remaining two \hx sources do not have obvious continuum
counterparts. The \hx spectra towards each of the four sources are shown in Fig.~\ref{specXnuc} and 
their line parameters are summarised in Table~\ref{tlineparaXnuc}. 
The sum of the \hx emissions from the four sources is 
(18$\pm$1)$\times$10$^{-23}$~W~m$^{-2}$, 
comparable to the total emission of 
(20~$\pm$~2)$\times$10$^{-23}$~W~m$^{-2}$ 
from the entire region.
Hence all of the extended emission observed within the 15~pc region can indeed be accounted for in terms of
four individual compact sources. 

\begin{table*}
\begin{center}
\caption[]{\bf Parameters for the \hx sources in the 15~pc region}
\label{tlineparaXnuc}
\begin{tabular}{lcccccccc}
\hline
\multicolumn{1}{l}{Source} & \multicolumn{1}{c}{Peak line} & \multicolumn{1}{c}{Central} & 
\multicolumn{1}{c}{FWHM$^b$} & \multicolumn{2}{c}{Integrated flux} & \multicolumn{2}{c}{Continuum$^c$} & \multicolumn{1}{c}{Source$^d$} \\
\multicolumn{1}{l}{name} & \multicolumn{1}{c}{flux density} & \multicolumn{1}{c}{velocity$^a$} &
\multicolumn{1}{c}{ } & \multicolumn{2}{c}{($\times$10$^{-23}$ W m$^{-2}$)} & \multicolumn{2}{c}{(mJy/beam)} & \multicolumn{1}{c}{size} \\
\multicolumn{1}{l}{ } & \multicolumn{1}{c}{(mJy/beam)} & \multicolumn{1}{c}{(km \sss)} & 
\multicolumn{1}{c}{(km \sss)} & \multicolumn{1}{c}{\hx} & \multicolumn{1}{c}{\hu} & 
\multicolumn{1}{c}{8.3 GHz} & \multicolumn{1}{c}{15 GHz} & \\
\hline
source 1$^e$ & 0.5  $\pm$ 0.1  & $\sim$220    & $\sim$150    & $\sim$2 & $<$9.1 & 10.5 & 9.5 & unres. \\
source 2     & 0.6  $\pm$ 0.1  & 180 $\pm$ 10 & 220 $\pm$ 20 & 3.9 $\pm$ 0.4 & $<$9.1 & 30 & 26 & unres. \\
source 3     & 0.75 $\pm$ 0.06 & 200 $\pm$ 10 & 240 $\pm$ 22 & 5.1 $\pm$ 0.9 & 20 $\pm$ 3 & 34 & 35& unres.  \\
source 4     & 1.15 $\pm$ 0.06 & 192 $\pm$ 6  & 198 $\pm$ 13 & 6.7 $\pm$ 0.5 & 23 $\pm$ 3 & 64 & 59& unres.  \\
\hline
\multicolumn{8}{p{5in}}{$^a$ \scriptsize Heliocentric velocity, optical definition.}\\
\multicolumn{8}{p{5in}}{$^b$ \scriptsize Corrections for finite velocity resolution not applied.}\\
\multicolumn{8}{p{5in}}{$^c$ \scriptsize Continuum flux per synthesized beam measured at the position of the line emission.}\\
\multicolumn{8}{p{5in}}{$^d$ \scriptsize Spatially unresolved, angular size $\leq$0.35\asecc.}\\
\multicolumn{8}{p{5in}}{$^e$ \scriptsize The \hx parameters of source 1 are less certain.}\\
\end{tabular}
\end{center}
\end{table*}

\begin{figure}
 \includegraphics[width=3.4in]{fig253_15.ps} \\
 \includegraphics[width=3.4in]{fig253_16.ps}
 \caption{
{\bf (a).} High resolution integrated VLA \hx line emission from the 15~pc region, summed over
velocities 316.6 to 90.8 km \sss. 
The synthesized beam is 0.38\asecc$\times$0.20\asecc, P.A.=$-$2$^{\circ}$. The contour levels are (1, 1.4, 1.8,
2.2, 2.6, 3. 3.4, 3.8, 4.2, 4.6, 5) times 10$^3$ Jy/beam$\times$Hz. The `star' symbols mark the
derived positions of the four line sources, as listed in Table~\ref{tlineXnucposn}. 
{\bf (b).} The positions of the four \hx sources are marked as `star symbols' on the contours of
15~GHz continuum emission imaged at high resolution. The 15~GHz continuum image is aligned using the
continuum and line emission from the radio nucleus. The synthesized beam of the continuum image is 
0.22\asecc$\times$0.10\asecc, P.A.=$-$15$^{\circ}$. }
\label{mom0nuc}
\end{figure}

\citet{ag90} imaged the \hx emission at a resolution of 7.4\asecc$\times$5.2\asec and 
detected a total line emission of 
44$\times$10$^{-23}$~W~m$^{-2}$, 
whereas our observations account for only 
25.5$\times$10$^{-23}$ W m$^{-2}$
over the entire disk. 
The remainder must be spatially extended so as to fall below the surface brightness 
sensitivity of the current observations. In this section, we have decomposed the observed \hx emission into eight distinct
compact sources. The positional offsets between these line sources and the associated continuum sources are 
typically $\leq$0.1\asecc. These offsets are therefore significant when compared to the size of the synthesized 
beam. The continuum emission in NGC 253 is dominated by non-thermal
radiation, whereas RRLs arise in thermal gas. Hence a spatial offset between the two emission regions could 
well exist, as in the case of NGC 1808 at high resolution \citep{koti96}. These offsets are probably due to lower
surface brightness of thermal compared to nonthermal emission, coupled with source confusion due to the 
high surface density of compact sources. Continuum emission imaged at a resolution higher than that of the 
current images should be able to delineate the actual continuum counterparts of the RRL sources. 

\section{A multi-density model for the ionised gas in NGC 253} 

RRL data can be used, along with the associated radio continuum measurements, to model the
physical properties of the line emitting (photo-)ionised gas. The atomic level populations are 
derived\footnote{The departure coefficients are taken from a program by \citet{salembro79}, 
modified by \citet{ww82} and \citet{pae94}.} 
assuming they are not in
local thermodynamic equilibrium (LTE). The ionized gas is modelled either as a uniform sphere,
a slab or a collection of spherical HII regions, all of constant density. 
Density, electron temperature, size (radius for spherical models and lateral size and thickness for slab
models) and the number of HII regions (for the collection of HII regions model) are free parameters. 
The temperature of the photo-ionized gas in all the models is taken to be 7500 K, a value typical 
of galactic HII regions, since
the results of the models do not depend significantly on the temperature. 
The values for the 
continuum and line flux densities are calculated for a grid of values of the free parameters and 
the observed flux densities are used to constrain the solution space. The range of derived values 
of the free parameters is determined primarily by the observational errors in the flux densities.
Externally stimulated emission 
is also considered (in addition to internally stimulated emission, see \citealt{azgv93}) using the
observed value of the continuum flux density. It is found that the derived range of the free parameters 
allow for substantial externally stimulated emission as well, but the values of these parameters are
not affected significantly due to limits to the continuum brightness temperature. Further details
of the modelling procedure have been described in \citet{mohan01}, and references therein.
In Sect.~\ref{sec_lowresmodel}, we attempt to model the entire low-resolution multi-frequency RRL data
as arising from a single component of ionised gas with uniform properties. 
In Sect.~\ref{sec_hydenmodel}, we use our higher 
resolution observations of cm-wave RRLs to derive the properties of the individual line emitting regions,
and finally, in Sect. 4.3, we propose a multi-density model for the ionised gas component in NGC 253.

\subsection{Modelling the low-resolution RRL data}
\label{sec_lowresmodel}

\begin{figure}
\resizebox{\hsize}{!}{\includegraphics{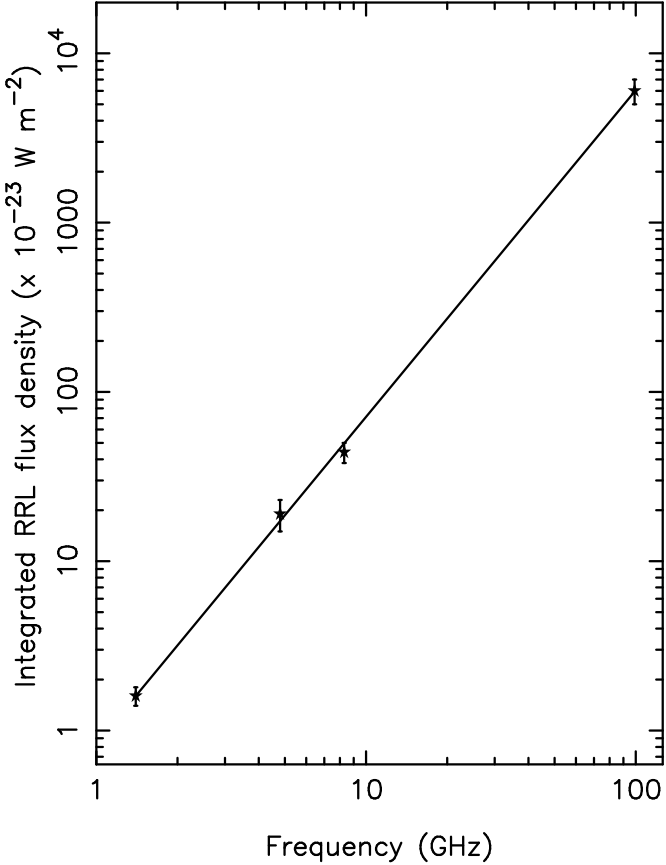}}
 \caption{Integrated line flux density $vs.$ frequency for NGC 253 for low resolution RRL data. The error bars 
are marked and the solid line is the best linear fit through the data. 1.4~GHz data : our work;
4.8 GHz and 8.3 GHz data : \citet{ag90};  99 GHz data : \citet{puxley97}.}
\label{rrlvsnu}
\end{figure}

\citet{puxley97} detected the 99~GHz H40$\alpha$ line in NGC 253 using single-dish observations and also
compiled low resolution interferometric RRL observations at 1.4~GHz, 4.8~GHz and 8.3~GHz \citep{ag90}. 
They show that the RRL data from 1.4~GHz up to 99~GHz is well fit by a F$_{\nu}$~$\sim$~$\nu^{2}$ curve 
(that is, the line spectral index $\alpha$=2) . In fact, our revised estimate of the 1.4~GHz line
flux density lies much closer to the best fit line than the earlier measurement. These data are plotted
in Fig.~\ref{rrlvsnu}. A function of the form F$_{\nu}$~$\propto$~$\nu^\alpha$ is 
fit to the data and the best fit value of $\alpha$ is 1.93 $\pm$ 0.04. 
The theoretically expected dependence of F$_{\nu}$ on frequency
when $\tau_c$$\ll$1, $|\tau_l|$$\ll$1, assuming LTE conditions and neglecting pressure
broadening, is indeed F$_{\nu}$~$\sim$~$\nu^2$, as noted by \citet{puxley97}.
They therefore suggest that all of the RRL emission from 1.4~GHz up to 99~GHz 
arises in LTE gas which is optically thin. This result is surprising since we do not expect 
either LTE conditions or optically thin conditions to hold for ionised gas over this range of frequencies. 

\begin{figure}
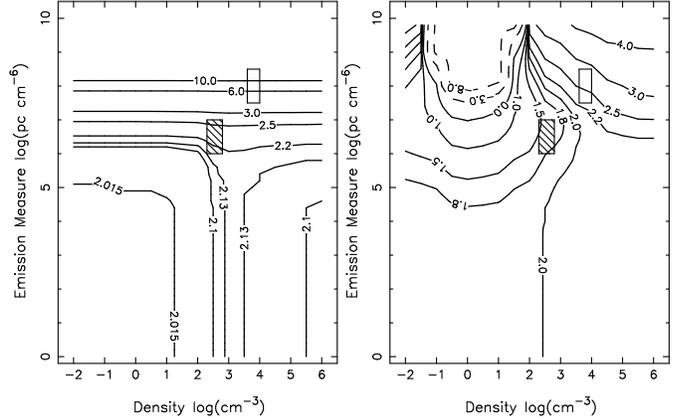

\resizebox{\hsize}{!}{\includegraphics{fig253_22.ps} \includegraphics{fig253_23.ps}}
 \caption{The expected dependence of the line spectral index $\alpha$ (defined as 
F$_{\nu}$~$\propto$~$\nu^{\alpha}$, where F$_{\nu}$ is the integrated flux density 
of the RRL at frequency $\nu$) between the 1.4 GHz H166$\alpha$ and the 99~GHz H40$\alpha$ 
line flux densities as a function of density and emission measure under {\bf (a).} LTE conditions 
and {\bf (b).} non-LTE conditions. The boxes mark the solution space for Components I (open 
box) and II (hatched box) of the ionised gas, derived in Sects.\ref{sec_hydenmodel} and 
\ref{sec_multimodel}.}
\label{alpharrl}
\end{figure}

In Fig.~\ref{alpharrl} we show the theoretical contours of the spectral
index $\alpha$ between 1.4~GHz and 99~GHz as a function of electron density and emission measure EM, 
including pressure broadening, for LTE as well as for non-LTE conditions. In the LTE case, 
$\alpha$ is quite close to 2.0 for EM$<$10$^6$ pc cm$^{-6}$ (or $\tau_c$ at 1.4~GHz $<$0.4)
and increases for higher values of EM due to optical depth effects. The weak dependence of 
$\alpha$ on density for low values of EM is due to pressure broadening. However, the behaviour of 
$\alpha$ is very different for the non-LTE case. The value of $\alpha$ 
is closer to 2.0 for certain high-EM regions in the density-EM plane compared to the LTE case. 
The ratio between the non-LTE and LTE flux densities is 1/b$_{\rm n}$ (b$_{\rm n}$ is the departure 
coefficient for level $n$), and is hence close to unity,
only for optically thin regimes. Hence, an LTE analysis of the line emission, as performed by
\citet{puxley97}, based on the observed $\nu^2$ dependence, will not be valid if the optical depth
is $\geq$0.5.

Therefore we investigate whether the observed line flux densities and the $\nu^2$ dependence 
of the integrated RRL flux from NGC 253 can indeed be modelled as emission arising from 
a single parcel of gas. 
The ionised gas is modelled as a rectangular slab at T$_{\rm e}$=5000 K, with the density \nee, 
lateral extent and thickness being the free parameters. 
The gas is required to produce the observed integrated line flux densities at the four frequencies.
The valid solutions obtained violate two observational constraints. The minimum lateral angular 
extent ($\sim$10\asecc) is larger than the observed value ($\lsim$6\asecc). Additionally, the 
predicted thermal continuum at 99~GHz is much higher than the estimated value from the observations 
of \citet{carl90} and \citet{peng96}. In a model where the calculated flux densities are normalised,
and only the calculated line spectral index $\alpha$ is constrained to be consistent with the
observed value of $\alpha$=1.93 $\pm$ 0.04, no solutions are obtained for LTE emission within 
2$\sigma$. The acceptable non-LTE models in this case are confined to \nee$\leq$100~\cc and 
EM$\leq$10$^4$ with some isolated high density solutions as well. An observed spectral index of
two, therefore, cannot be used to infer emission measures assuming LTE conditions.

Hence, all of the RRL emission from 1.4~GHz up to 99~GHz cannot arise in the same component of ionised gas. 
However, the morphology of the line emission in the \hx line images is observed to be complex and consists of
individual sources, which may have different properties (for example, the line from 
the radio nucleus arises in gas at a density 10 times higher than derived in the above models and 
contributes about 10 \% of the total ionisation derived; \citealt{mohan02}). Therefore each of 
the compact line emitting sources must be modelled separately, as is done in the following section.

\subsection{Modelling the cm-wave RRLs: High density ionised gas}
\label{sec_hydenmodel}

The physical properties of the compact \hx emitting sources can be modelled in a more realistic
fashion with our high resolution data since the smaller synthesized beam size strongly constrains the 
projected lateral size of the ionised gas. 
The four extranuclear line sources and the four sources inside the central 15 pc region 
are modelled individually. The models are constrained to produce the observed
\hx and \hu flux densities. The free-free emission at 8.3~GHz and 15~GHz 
is also constrained to be less than the measured continuum flux densities. Further, the angular 
size of the gas in these spatially unresolved sources is constrained to be $<$0.35\asecc, corresponding
to the size of the synthesized beam.
The observational constraints used in the modelling are summarised in Table~\ref{talllinepara}
for the extranuclear sources and in Table~\ref{tlineparaXnuc} for the sources in the 15 pc region.
The valid solutions obtained for the extranuclear sources are listed in Table~\ref{tmodelXxnuc} and 
notes on each object are given below.

\begin{table*}
\begin{center}
\caption[]{\bf Model results for the extranuclear \hx sources}
\vspace{0.5cm}
\label{tmodelXxnuc}
\begin{tabular}{lcccc}
\hline
\multicolumn{1}{l}{Parameter} & \multicolumn{1}{c}{5.49$-$42.3} & \multicolumn{1}{c}{5.59$-$41.6} & 
\multicolumn{1}{c}{5.72$-$40.1} & \multicolumn{1}{c}{5.75$-$39.9} \\
\hline
Density ($\times$10$^3$ \ccc)        &  4-10 & 2-20 & 2-10 & 2-10 \\
Effective size (pc)                  &  2-3.5$^a$ & 1-5 & 1.4-2.7 & 1.7-2.9 \\
$\tau_{\rm c}$ at 8.3 GHz            &  0.2-1.5 & 0.2-3 & 0.2-1.2 & 0.1-1.5 \\ 
S$_{\rm ff}$ at 8.3 GHz (mJy)        & 0.5-0.7 & 1-2.5 & 2-3 & 2-3 \\
\nlyc ($\times$10$^{51}$ ph. \sss)   & 1  & 2 & 1-10 & 2-5 \\
Mass ($\times$10$^3$ \msunn)         &  0.16-0.36 & 0.3-1 & 0.35-1.3 & 2 \\
4.8 GHz free-free flux density (mJy) & 0.2-0.4 & 0.7-2 & 1-4 & 1-2.5 \\
99 GHz free-free flux density (mJy)  & 0.7-0.9  &  1.3-2  &  2-3.5  &  2-4 \\
1.4 GHz line flux ratio$^b$ (\%)      & 0  &  0.1 & $<$1 & 0.05 \\
4.8 GHz line flux ratio$^b$ (\%)      & 0.2-0.8  &  0.7-1.8  &  1-4  &  1-2.5 \\
99 GHz line flux ratio$^b$ (\%)       & 0.2-0.3  &  0.4-0.7  &  0.7  & 0.7 \\
\hline
\multicolumn{5}{p{5in}}{$^a$ \scriptsize Only thin slab solutions with thickness between 0.01-0.04 pc.}\\
\multicolumn{5}{p{5in}}{$^b$ \scriptsize Ratio of predicted RRL flux of the eight \hx sources to the observed value.}\\
\end{tabular}
\end{center}
\end{table*}

\begin{table*}
\begin{center}
\caption[]{\bf Model results for the \hx sources in the 15~pc region}
\label{tmodelXnuc}
\begin{tabular}{lcccccc}
\hline
\multicolumn{1}{l}{Parameter} & \multicolumn{2}{c}{Source 1$^a$} & \multicolumn{2}{c}{Source 2$^a$} & 
\multicolumn{1}{c}{Source 3} & \multicolumn{1}{c}{Source 4} \\
\cline{2-3}
\cline{4-5}
\multicolumn{1}{l}{} & \multicolumn{1}{c}{Low} & \multicolumn{1}{c}{High} & \multicolumn{1}{c}{Low} & 
\multicolumn{1}{l}{High} & \multicolumn{1}{c}{} &\multicolumn{1}{c}{} \\
\hline
Density ($\times$10$^3$ \ccc)  & 2 & 10 & 0.9 & 6 & 8-10 & 7-10 \\
Effective size (pc) & 5 & 1.8 & 6.4 & 2.6-3.7 & 2-2.5 & 2.5 \\
$\tau_{\rm c}$ at 8.3 GHz & 0.2 & 1.7 & 0.05 & 0.8 & 1-2.5 & 1-2 \\
S$_{\rm ff}$ at 8.3 GHz (mJy)  & 0.4 & 0.3 & 2 & 4 & 6 & 5.5 \\
\nlyc ($\times$10$^{51}$ ph. \sss) & 2.5-5 & 4-5 & 1.5 & 4 & 8-15 & 6-14 \\
Mass ($\times$10$^3$ \msunn) & 3-5 & 0.8 & 3.2 & 1.3 & 1.5-2.5 & 1.3-2 \\
4.8 GHz free-free flux density (mJy) & 3-5  & 1.6 & 1.2 & 3-5 & 2.7 & 2.5-5 \\
99 GHz free-free flux density (mJy) & 2.3-4.3 & 4 & 1.8 & 2-7 & 6-10 & 6-12 \\
1.4 GHz line flux ratio$^b$ (\%)& 1 & 0.05 & 19  & 0.1 & 0.05 & 0.05 \\
4.8 GHz line flux ratio$^b$ (\%) & 5 & 1.6 & 13 & 5 & 3.5 & 5 \\
99 GHz line flux ratio$^b$ (\%)& 0.8 & 1.1 & 0.3 & 1 & 2.5 & 2.5 \\
\hline
\multicolumn{7}{p{5in}}{$^a$ \scriptsize For sources 1 and 2, the density range is large;
typical low and high density solutions are listed.}\\
\multicolumn{7}{p{5in}}{$^b$ \scriptsize Ratio of predicted RRL flux of the eight \hx sources to the observed value.}\\
\end{tabular}
\end{center}
\end{table*}

\noindent (1) {\em 5.49$-$42.3:} No solutions are obtained for a spherical HII region
model; uniform slab models yield solutions corresponding to thin slabs. Since this source is 
not listed as a \ha emitter by \citet{forbes00}, the integrated \ha flux of 2$\times$10$^{-15}$ W m$^{-2}$
derived from the ionisation rate implies a (foreground) extinction of \av~$>$~10. \\
(2) {\em 5.59$-$41.6 (= IR peak ?):} \citet{pina92}, \citet{sams94} and others 
have shown that the radio nucleus and the IR peak are two separate objects, and 
that astrometric errors do not allow an unambiguous association of any other radio 
source with the IR peak. The closest radio sources are 5.59$-$41.6 and 5.62$-$41.3, 
which have spectral indices of +0.61 $\pm$ 0.15 and $-$0.21 $\pm$ 0.14 respectively, 
and are hence dominated by thermal gas. However, the detection of recombination lines from 
5.59$-$41.6 leads us to suggest that this is indeed the IR peak. This issue is discussed further 
in Sect.~\ref{sec_irpeak}.\\
(3) {\em 5.72$-$40.1:} We model the line emission from this thermal object along 
with the continuum constraints derived by UA97\nocite{ua97}. The derived gas mass, 
density, and size are similar to the UA97 values. However, only an insignificant fraction 
of the continuum at $\nu$$\leq$8.3~GHz is thermal, whereas the continuum spectral 
index is observed to be flat at frequencies $\geq$~4.8~GHz. Hence, a single-density model cannot explain the 
observed flat spectral index as well as the RRL emission. Using the estimated 
\ha strength of 1.2$\times$10$^{-14}$ W m$^{-2}$ and identifying the source with 
a \ha knot in \citet{forbes00} $\sim$0.4\asec eastwards, the (foreground) 
extinction \av is estimated to be~13. \\
(4) {\em 5.75$-$39.9:} The line emitting gas is modelled assuming that the 
free-free flux density of the thermal gas does not exceed the measured continuum 
of 5.5 mJy at 8.3 GHz. No \ha or IR emission knots are observed in this region \citep{forbes00} and hence 
the extinction \av must be $>$~9.

The solutions obtained for the sources in the central 15 pc region are similar 
for a uniform sphere as well as a rectangular slab geometry, and are described 
in Table~\ref{tmodelXnuc}. The derived range of densities for sources 1 and 2 
are wider than those for sources 3 and 4 since only upper limits to 
the \hu line flux density are available for the former two. The spectral 
indices of RRL flux density between pairs of frequencies from 1.4~GHz to 99~GHz
derived from the models range between 0.5 and 5.0 which differ significantly
from 2.0 (see Sect.\ref{sec_lowresmodel}). The derived properties of 
the four sources (Table~\ref{tmodelXnuc}) are similar. 

The gas densities of the eight compact \hx sources lie in the range 
5$\times$10$^3$-10$^4$~\ccc, similar to compact HII regions in our galaxy. 
To illustrate the advantage of high resolution observations (which set stricter constraints
on the projected size), we also model the 15~pc region as a whole. The results obtained are
different compared to those derived above (lower densities, larger sizes and a dominant 
contribution from stimulated emission). However, the results are similar to those derived 
by \citet{puxley97} from low resolution data. Hence, high resolution observations are 
essential for reliably determining the properties of the compact component of ionised gas.

\subsection{Modelling the decimetre-wave RRLs: multi-density medium}
\label{sec_multimodel}

The sum of the observed continuum and line flux densities of the eight \hx sources, derived 
in Sect.~\ref{sec_hydenmodel}, are listed in Table~\ref{tcomponents} (as Component~I). 
This component contributes no more than 2~\% (or 21~\% for the low density 
models) of the observed H166$\alpha$ emission at 1.4~GHz, between 18-36 \% of 
the H110$\alpha$ emission at 4.8~GHz and about 9 \% of the H40$\alpha$ 
emission at 99~GHz. Therefore, in order to explain the remaining RRL 
emission at these frequencies, additional components of ionised gas, 
different from that detected in our high resolution \hx data, must exist.
No plausible models can account for the remainder of the 1.4~GHz as well as the 99~GHz 
line emissions to arise in the same component of gas. Ionised gas at densities
$>$10$^4$ \cc is needed to account for the observed mm-RRL emission, and
this gas would produce negligible line emission at 1.4~GHz due to continuum optical depth 
effects. Hence, a minimum of three components of ionised gas seem to be needed 
to explain all of the RRL data on NGC 253 : Component I which produces the high 
brightness part of the \hx emission, described in the previous section, Component II
which explains most of the 1.4~GHz line and the remainder of the 4.8~GHz 
and 8.3~GHz lines, and Component III, which leads to the mm-wave line emission.

\begin{table*}
\begin{center}
\caption[]{\bf Observational and derived parameters of Components I, II \& III} 
\label{tcomponents}
\begin{tabular}{lccc}
\hline
\multicolumn{1}{l}{Observational parameter} & \multicolumn{1}{c}{Component I} & \multicolumn{1}{c}{Observations} & 
\multicolumn{1}{c}{Comp.s II+III} \\
\multicolumn{1}{l}{ } & \multicolumn{1}{c}{} & \multicolumn{1}{c}{(low res.)} & 
\multicolumn{1}{c}{(expected)} \\
\hline
H166$\alpha$ at 1.4 GHz ($\times$10$^{-23}$ W m$^{-2}$) & 0.03-0.3 & 1.6 & 1.5 $\pm$ 0.4 \\
H110$\alpha$ at 4.8 GHz ($\times$10$^{-23}$ W m$^{-2}$) & 3.4-6.9 & 19.1 & 14.5 $\pm$ 4 \\
H92$\alpha$ at 8.3 GHz ($\times$10$^{-23}$ W m$^{-2}$) & 25.5 & 44.4 & 19 $\pm$ 3 \\
H40$\alpha$ at 99 GHz ($\times$10$^{-23}$ W m$^{-2}$) & 564 & 6270 & 5706 \\
1.4 GHz continuum (mJy) & 10 & 852 & $<$842 \\
4.8 GHz continuum (mJy) & 11-27 & 1000 & $<$990 \\
8.3 GHz continuum (mJy) & 25 & 800 & $<$775 \\
99 GHz continuum (mJy) & 23-42 & 75 & $<$52 \\
\hline
\multicolumn{1}{l}{Derived parameter} & \multicolumn{1}{c}{Component I} & \multicolumn{1}{c}{} &
\multicolumn{1}{c}{Component II} \\
\hline
Density (\ccc) & $\sim$10$^4$ & & 300-1000   \\
Size (pc)      & $<$10 & & 30-60   \\
\nlyc ($\times$10$^{52}$\sss)   & 4 & & 3-6  \\
\hline
\multicolumn{4}{p{5in}}{\scriptsize  Col (1): Parameter; col (2): predicted value of the parameter, summed
over the eight compact \hx sources (Component I, see Sect.~\ref{sec_hydenmodel}); 
col (3): observed value of parameter, from low resolution data; 
col (4): the difference between col. 3 and col. 2, which is modelled as Components II and III in 
Sect.~\ref{sec_multimodel}.}
\end{tabular}
\end{center}
\end{table*}

Component II is modelled to produce all of the remaining 1.4~GHz, 4.8~GHz and 8.3~GHz
line emissions, with the 99~GHz line flux density imposed as an upper limit. The transverse
size is constrained to be $<$7.5\asecc, the size of the low resolution emissions.
The derived electron densities are between 300-1000~\ccc, effective 
sizes between 30-60~pc and ionising photon rate is (3-6)$\times$10$^{52}$~\sss. 
The stimulated emission component due to the background nonthermal radiation is 
between 50-90~\%. If the stimulated emission is
set to zero, we obtain solutions for densities between 200-500~\ccc, but with very similar 
ionisation rates. The derived properties of the gas in Components I and II are 
listed in Table~\ref{tcomponents}.
This gas (Component II) contributes 10-12~\% of 
the remaining 99~GHz emission. 
Hence a third higher density component is obviously necessary. 
Since the H40$\alpha$ line was detected by \citet{puxley97} at low signal to noise, we 
do not attempt to model this high frequency line emission in detail. 

\section{Discussion}

Though high resolution radio continuum observations of NGC 253 exists, recombination lines
provide additional advantages. The local gas density can be derived from RRLs whereas 
continuum data yield only the rms density, and the densities derived in this work are higher
than the rms densities for some of the thermal sources. Moreover, the thermal fraction of 
the continuum emission is not known accurately due to the large errors in the observed 
continuum spectral indices. Such errors also result to difficulties in determining free-free
optical depth effects, especially in the case of multiple density components inside a single 
source.

The continuum spectral index measurements by UA97 indicate that thermal gas is present over a
substantial area and also forms a significant part of the continuum emission from a number of
compact sources. Hence, it is notable that 8.3~GHz and 15~GHz line emission is detected from 
only a handful of sources, which may be ascribed partly to lack of adequate sensitivity. 
Additionally, since the observed RRLs are sensitive primarily to gas at densities 
10$^3$-10$^4$ \ccc, the line emission would not be traced efficiently by 
high resolution observations of the \hx line emission
if a substantial part of the thermal gas detected in the continuum is at very different densities. 
Lower density diffuse gas, for example, can give rise to cm-wave RRL emission which is below the 
sensitivity limit of our observations while still being able to make the observed continuum 
spectral index flatter, thus explaining the presence of thermal sources not detected in RRLs
in our observations. 

\subsection{The IR peak}
\label{sec_irpeak}

The IR peak dominates the NIR and MIR continuum, \brg and [NeII] lines and accounts for 10~\% of 
the total bolometric IRAS flux from NGC 253, and is hence a prominent thermal source. 
However, this source is not conspicuous in the radio continuum image which is 
understandable since the total radio continuum is predominantly non-thermal. \brg data from 
\citet{forbes93} implies a free-free flux density in the radio of only $\sim$1 mJy in the 
for the IR peak for optically 
thin gas (and 2.5 mJy for \avv=10); observations of \citet{kww94} are consistent if a large 
MIR excess is present. However, the IR peak does not have a strong RRL counterpart, which is surprising. 
Much higher extinction towards
the central 15 pc region than towards the IR peak could be invoked. Nevertheless, since the 
12.8~$\mu$m [NeII] emission is much stronger towards the IR peak than towards the centre, 
the required extinction towards the latter would need to be unrealistically high, unless most of the 
Ne in the centre is doubly ionised due to the AGN. However, low density models can be 
constructed which can explain the weak RRL emission along with a high star formation rate. Hence we
suggest that the thermal gas in the IR peak is probably at densities lower than $\sim$few 100 \cc and is 
therefore not prominent in high frequency line observations. We predict that a high resolution 
observation of the H166$\alpha$ line using the VLA in the A configuration would show the IR 
peak to be a prominent line source since this line is more sensitive to low density gas and the VLA in
the A configuration will be able to resolve the IR peak from the surrounding emission.

\subsection{Ionisation sources of cm-RRL emitting gas}

It was argued by \citet{mohan02} that the RRL emitting gas in the radio nucleus 
is ionised by a weak central AGN, for which additional evidence was presented by 
\citet{weaver02}. The nature of ionisation of the remainder of \hx emission is
discussed in this section.

If the individual \hx RRL sources in the 15~pc region are ionised by stars, the
derived values of the ionisation rates and sizes imply the presence of a Super Star 
Cluster (SSC) within each (based on the typical sizes and luminosities of SSCs derived
by \citealt{meurer95}). Similarly, massive clusters or SSCs would be required for the 
four sources outside the central region. However, the radio recombination line widths 
(presumably caused by the expansion of the HII regions) are supersonic, $\sim$200 km~\s 
for the central sources and 70-100 km \s for the latter ones. Hence the dynamical age for 
the ionised gas is $\sim$10$^4$ years for all of the sources in the central 15~pc region and 
between (4-10)$\times$10$^4$ years for the extranuclear sources. These are too young to be realistic
since it is improbable that star clusters much younger than 10$^5$ years would have formed 
enough OB stars to provide an ionisation in excess of 10$^{51}$ photons~\sss. The thermal gas in
the four nuclear sources is not gravitationally bound to the clusters within, though gravity could
play a role in the dynamics of the gas in the extranuclear sources. This ``lifetime problem" is 
similar to the situation in NGC 5253 and He 2-10 \citep{mohan01}. The actual age of the 
objects could be increased by invoking internal mass loading whereby there is injection of 
mass into the ionised component from evaporation of neutral or molecular clumps. This can
increase the density of gas within and maintain a constant Str\"omgren radius 
\citep{dwr95,lizano96}. 
There could also be a substantial contribution from an 
unordered or turbulent component to the observed line width, in which case the dynamical
time of the ionized gas will be longer compared to an ordered outflow. 
However, since the central 15~pc region is also deficient in PAH and MIR emission compared
to the IR peak \citep{keto99}, the possibility of ionisation of these HII regions 
solely by SSCs is not entirely satisfactory. A possible explanation for the deficiency
of nuclear PAH emission is dissociation by the hard radiation from the central AGN
\citep{keto99}. 

\citet{weaver02} has shown that the radio nucleus is powered by a weakly accreting 
supermassive black hole or an intermediate mass black hole. Assuming the former, 
it is possible that the kinematics of the gas in the central 15~pcs is influenced by 
the gravitational potential of the central mass. Hence a fraction of the large observed
line widths could be explained by the AGN, though the remainder would still remain 
supersonic. The gas could also be partially ionised by the AGN, though star clusters
would have to provide all of ionisation for the extranuclear sources. There is evidence
that the gas kinematics in the central region of NGC 253 is complex (see \citealt{day01}),
and involves multiple orbits at different orientations seen at high inclination. Hence it is
possible that the RRL sources in the 15~pc region may not contain individual star
clusters, but could be patterns or enhancements of emission measure along the line of 
sight. In such a case,
the stellar densities needed for ionisation can be much lower, and the line widths 
of individual ionized gas components can also be
smaller than the observed value, averting the ``lifetime problem".
However, the derived properties of the line emitting gas would remain unchanged.
We predict that with increased spatial resolution, the line widths 
of the central RRL sources would decrease, similar to the [NeII] and RRL line widths in the 
Galactic Centre. In fact, due to the presence of a weak AGN, star clusters, high density 
ionised gas, and complex kinematics, the centre of NGC 253 is qualitatively similar 
to the Galactic Centre.

\subsection{Multi-density model of the ionised gas}

Modelling the multi-frequency high resolution RRL data on NGC 253 led to a multi-density 
model for the ionised gas, similar to that in Arp 220 \citep{anan00}. In contrast,
using the low resolution RRL data at precisely the frequencies we have modelled here, 
\citet{puxley97} could model all of the RRL emissions as arising from the same
component of gas. The solution space of Components I and II in the \nee-EM plane are plotted
in Fig.~\ref{alpharrl}. The figure for non-LTE conditions shows that the line spectral indices
between 1.4~GHz and 99~GHz is not 2.0 for either component, though the spectral index of the total
line emission is indeed very close to 2.0. Thus, high spatial resolution, coupled with 
multi-frequency data, is necessary to adequately model the compact ionised gas component.

High resolution observations of molecular gas in NGC 253 within the central few hundred parsecs 
(HCN: \citealt{paglione95}; CS: \citealt{peng96}) indicate that the gas density exceeds 
10$^4$ \ccc. Hence Component I of ionised gas could well be produced by ionising this 
surrounding molecular material. Since the ionised gas becomes less dense and also larger in size
with age, due to expansion of the HII regions, it is consistent to expect the more compact and denser
Component I to have similar density as the molecular gas.
The spectral index of the global radio continuum emission in the 
central 25~pc of NGC 253 is not steep ($\sim$$-$0.5; UA97, this work). From the spectral index of the continuum 
emission and also based on the [NeII] emission, we can conclude that free-free emission 
is not, however, the dominant portion of the total continuum emission. Therefore, the presence of
foreground ionised gas with large covering factor (which absorbs the background non-thermal
emission) is implied. Component II does satisfy these requirements, since it has a large transverse 
size and has sufficient optical depth at 8~GHz. However, the exact location of this 
component of ionised gas needs to be investigated with future VLA A configuration observations
of the 1.4~GHz RRL. Though the properties of Component III have not been derived here,
this component can be modelled in detail using the observations of the 45~GHz RRL and 
also mm-wave RRLs, which have been carried out using the VLA and IRAM respectively (Zhao, 
and Viallefond, private communication).

In this work, a three-component model of the ionised gas has been derived using observations 
of RRLs at three different frequencies. This raises the question of uniqueness of the decomposition
into these specific density components. The ionised gas in NGC 253 is not expected to exist 
at three discrete densities; we interpret these results as an approximation to the intrinsic 
density distribution of the gas. This intrinsic distribution, presumably continuous, is a 
convolution of the three effects, namely, the initial density structure of the molecular 
ISM, the detailed time dependence of the star formation rate, and the dynamics of the HII 
regions. By observing widely spaced RRL, each of which is sensitive to a different range of densities, 
the properties of the ionised gas can be approximated in terms of a finite number of discrete density ranges.
Further, the sizes and masses of the gas in each of these density ranges can also be predicted. 
Hence, we divide the ionised gas into three density bins, the exact ranges of which are decided 
partly by the frequencies of the observed RRLs. We can then accurately determine the
length scales and masses and hence the required ionisation rates for the gas in each of these bins by
modelling the line formation. Future observations of a larger number of RRLs from decimetre to 
millimetre may be used to constrain a more general density distribution function directly.

\section{Summary and Conclusions}

The 1.4~GHz line emission is detected with a resolution of $\sim$50 
pc. The 8.3~GHz \hx and 15~GHz \hu line emissions have also been imaged, at high resolutions ($\sim$4 pc and 
$\sim$2 pc respectively). The total \hx emission, which is about half the value measured at low resolution, consists 
of an extended region of emission surrounding the radio nucleus (the `15~pc region') and four compact sources in 
the outer disk of NGC 253. The line emission in the 15~pc region can be decomposed into four additional compact 
sources. The peak of the line emission is towards the radio nucleus. The line widths of the outer 
sources is $\leq$100 km~\s whereas the line widths of the sources in the inner 15~pc are 
about 200 km~\sss. \hu emission has been detected
from only two of these sources. Not all the \hx line sources have identifiable continuum counterparts. 

In contrast to existing interpretations, we show that multi-wavelength RRL emission from NGC 253 does not 
arise in optically thin gas in LTE. The four extranuclear \hx sources as well as the four compact sources 
in the 15~pc region are modelled individually and arise in gas at densities close to 
10$^4$ \ccc. The total ionisation and gas mass in this high density Component~I are 
$\sim$4$\times$10$^{52}$ \s and 5000-10000 \msunn. The total size subtended by Component~I is $<$10 pc. 
This component produces a negligible fraction of the 1.4~GHz line emission. 

A second component of ionised gas, Component~II, is required to explain all of the observed H166$\alpha$ line 
at 1.4~GHz and the remainder of the cm-wave RRL emissions. This gas is a lower density of a few 100~\ccc. 
The The typical transverse size is a few tens of pc. A third component of gas, whose density must be higher 
than 10$^4$ \ccc, is also needed to explain the observed H40$\alpha$ emission at 99~GHz, but has not 
been modelled in this work.  In conclusion, 
\begin{itemize} 
\item The ionised gas component in NGC 253 is 
decomposed into three distinct components: a low density component ($\sim$500 \ccc) which contributes 
most of the observed thermal radio continuum emission and almost all of the 1.4~GHz RRL; a higher density 
component ($\sim$10$^4$ \ccc) which is more localised and exhibits supersonic line widths, and a third 
component at an even much higher density, responsible for mm-wave RRLs.  
\item The nature and location of the IR peak is uncertain and probably contains low density 
gas ionised by a star cluster which is much less obscured than the central 15~pc region.  
\item The ionisation of the line emitting high density thermal gas in the 15~pc region is probably determined
by obscured star clusters and the dynamics could be partly  determined by the AGN, like the Galactic Centre. Alternatively, 
some of the compact line sources could be line of sight enhancements in emission measure due to complex
dynamics in the region, rather than individual HII regions surrounding massive star clusters. In 
either case, we have discovered further examples of supernebulae with large line widths sustained by internal 
gas dynamical processes.
\end{itemize}

\begin{acknowledgements}
We acknowledge the referee for several useful comments and suggestions which have
improved the presentation of the paper and clarified certain issues better.
We thank C. L. Carilli for the use of his 330 MHz continuum image of NGC 253. 
We wish to acknowledge K. S. Dwarakanath, Biman Nath, B. Ramesh, K. Johnson, A. Omar, 
Rekesh Mohan, Shiv Sethi, Mousumi Das and P. Manoj for useful discussions 
and helpful comments.
The VLA is a facility of the National Radio Astronomy Observatory (NRAO), which is 
operated by Associated Universities, Inc. under a cooperative agreement with the 
National Science Foundation.
\end{acknowledgements}

\appendix 

\section{Radio continuum sources in the disk of NGC 253}

\begin{figure}
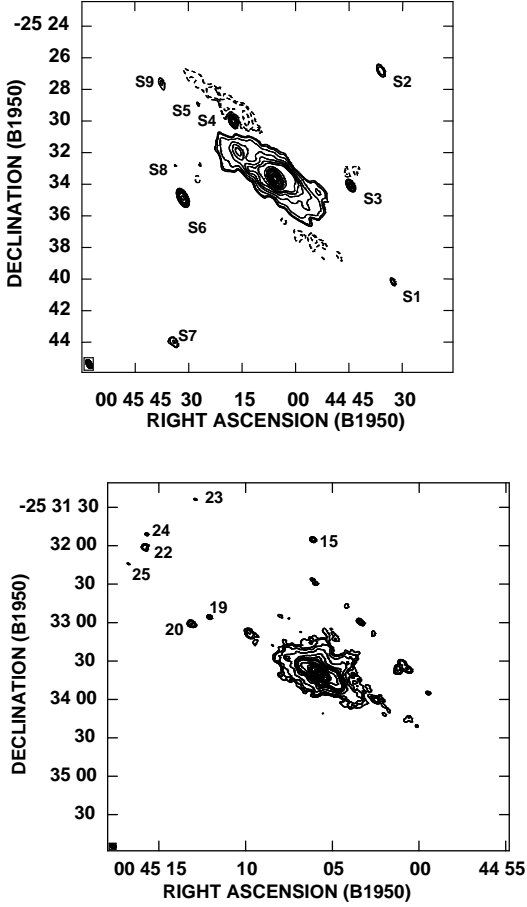

 \includegraphics[width=2.3in]{fig253_24.ps} $~~~~~~~$
 \includegraphics[width=2.6in]{fig253_25.ps}
 \caption{
{\bf (a)} The 1.4 GHz continuum emission from NGC 253 taken using the VLA C configuration
data, showing sources identified with the galaxy. The synthesized beam is 35\asecc$\times$17\asec at a 
P.A. of 26$^{\circ}$. The contour levels are in steps of (-20, -10,
-7, 7, 10, 20, 40, 60, 80, 100, 200, 400, 600, 800, 1000, 2000, 4000, 6000, 8000) times the rms, 
0.2 mJy/beam.
{\bf (b)} Higher resolution BnA configuration image. The area shows the sources detected by \citet{ulvi00}.
The synthesized beam is 3.8\asecc$\times$2.6\asec at a P.A. of 61$^{\circ}$. The rms
of the image is 0.1 mJy/beam and the contour levels are in steps of (5, 7, 10, 20, 40, 60, 80, 100,
200, 400, 600, 800, 1000, 2000, 3000, 4000) times the rms. Negative contours are below $-$5$\sigma$.}
\label{contCL}
\end{figure}

The 1.4 GHz H166$\alpha$ observations performed using the VLA in the C configuration, described 
in \citet{ag90}, were reanalysed. The 1.4 GHz RRL was detected, and was found to coincide with
the position of the peak of the radio continuum (radio nucleus). 
This dataset is used to image the continuum emission from NGC 253 at a higher sensitivity and over a
larger field of view than what is available in the literature.
Sources within 20\amin of the centre of NGC 253 which are within the spatial extent of 
the 330 MHz continuum emission \citep{carilli92,carilli96} are identified with the galaxy.
These are named as S1 to S9 and are marked in Fig.~\ref{contCL}a. \citet{ulvi00}
listed 22 circumnuclear sources within the inner 3\amin ($<$2 kpc) region. Using our C
as well as BnA configuration images, we mark those of their sources which have uncertain 
parameters, as well as new sources, in Fig.~\ref{tcontCL}b. The flux densities and positions 
of these continuum sources were determined using JMFIT in AIPS and the catalogue list is presented 
in Table~\ref{tcontCL}. 

\begin{table*}
\begin{center} 
\caption[]{\bf Continuum sources in the 1.4 GHz VLA C and BnA configuration images}
\label{tcontCL}
\begin{tabular}{lccccl}
\hline
\multicolumn{1}{l}{Source} & \multicolumn{2}{c}{Position (B1950.0)} & \multicolumn{1}{c}{S$_{1.4~\rm GHz}$} & 
\multicolumn{1}{c}{Source} & \multicolumn{1}{c}{Comments} \\
\multicolumn{1}{c}{name} & \multicolumn{1}{c}{$\alpha$ (00$^h$)} & \multicolumn{1}{c}{$\delta$ ($-$25$^{\circ}$)} & 
\multicolumn{1}{c}{(mJy)} & \multicolumn{1}{c}{size} & \multicolumn{1}{c}{}  \\
\hline
S1     & 44$^m$ 32$^s$.65 & 40\amin 11\asecc.2& 3.4  & U$^a$ & \\
S2     & 44$^m$ 35$^s$.95 & 26\amin 52\asecc.6& 10.8 & 4.0\asecc$\times$1.6\asec & \\
S3$^b$ & 44$^m$ 44$^s$.57 & 34\amin 06\asecc.0& 19.0 & 0.5\asecc$\times$0.4\asec & {\footnotesize B1} \\
S4$^b$ & 45$^m$ 17$^s$.56 & 30\amin 00\asecc.0& 25.4 &  & {\footnotesize B2; 3 point sources }\\
S5$^b$ & 45$^m$ 27$^s$.50 & 28\amin 53\asecc.0& 1.8  & U & {\footnotesize real feature ?} \\
S6     & 45$^m$ 31$^s$.57 & 34\amin 53\asecc.7& 56   & 2.4\asecc$\times$0.5\asec & {\footnotesize B3} \\
S7     & 45$^m$ 34$^s$.03 & 43\amin 59\asecc.7& 7.2  & 3\asecc$\times$1\asecc? & {\footnotesize possibly resolved } \\
S8     & 45$^m$ 33$^s$.79 & 32\amin 49\asecc.1& 1.6  & U &  \\
S9     & 45$^m$ 37$^s$.70 & 27\amin 34\asecc.9& 2.8  & U &      \\
\hline
15$^c$ & 45$^m$ 06$^s$.12 & 31\amin 55\asecc.3& 2.0  & 2\asecc$\times$2\asec & {\footnotesize possibly resolved} \\
19$^c$ & 45$^m$ 12$^s$.07 & 32\amin 55\asecc.7& 1.2  & 4.5\asecc$\times$2.1\asec & \\
20$^c$ & 45$^m$ 13$^s$.13 & 33\amin 01\asecc.1& 3.5  & 4.1\asecc$\times$3.4\asec & \\
22$^c$ & 45$^m$ 15$^s$.75 & 32\amin 01\asecc.2& 1.5  & 8.6\asecc$\times$5.6\asec & {\footnotesize within diffuse em.}\\
23$^d$ & 45$^m$ 12$^s$.91 & 31\amin 23\asecc.9& 0.6  &  --- & too weak\\
24$^d$ & 45$^m$ 15$^s$.69 & 31\amin 51\asecc.2& 0.9  &  --- & too weak\\
25$^d$ & 45$^m$ 16$^s$.80 & 32\amin 13\asecc.9& 0.7  &  --- & too weak\\
\hline
\multicolumn{6}{p{5in}}{$^a$ \scriptsize U - unresolved, even in the BnA configuration images.}\\
\multicolumn{6}{p{5in}}{$^b$ \scriptsize These are within 2-3 synthesized beams of $-$ve extended emission in
the C configuration image.}\\
\multicolumn{6}{p{5in}}{$^c$ \scriptsize Sources in U00 (\citealt{ulvi00}) with inaccurate flux densities, positions.}\\
\multicolumn{6}{p{5in}}{$^d$ \scriptsize Sources seen in BnA image but not in \citet{ulvi00}.}\\
\end{tabular}
\end{center}
\end{table*}

{}


\begin{thebibliography}{}
\bibitem[Anantharamaiah \& Goss(1990)]{ag90}Anantharamaiah, K. R., \& Goss, W. M. 1990, in IAU Colloq. 125, Radio Recombination Lines: 25 Years of Investigations, ed. M. A. Gordon \& R. L. Sorochenko (Dordrecht: Kluwer), 267
\bibitem[Anantharamaiah \& Goss(1996)]{ag96}Anantharamaiah, K. R. \& Goss, W. M. 1996, \apj, 466, L13
\bibitem[Anantharamaiah et al.(2000)]{anan00}Anantharamaiah, K. R., Viallefond, F., Mohan, N. R., Goss, W. M., \& Zhao, J. -H. 2000, \apj, 537, 613  
\bibitem[Anantharamaiah et al.(1993)]{azgv93}Anantharamaiah, K. R., Zhao, J. -H., Goss, W. M., \& Viallefond, F. 1993, \apj, 419, 585
\bibitem[Antonucci \& Ulvestad(1988)]{au88}Antonucci, R. R. J. \& Ulvestad, J. S. 1988, \apj, 330, L97
\bibitem[B\"oker et al.(1998)]{boker98}B\"oker, T., Krabbe, A., \& Storey, J. W. V. 1998, \apj, 498, L115
\bibitem[Carilli(1996)]{carilli96}Carilli, C. L. 1996, \aap, 305, 402
\bibitem[Carilli et al.(1992)]{carilli92}Carilli, C. L., Holdaway, M. A., Ho, P. T. P., \& de Pree, C. G. 1992, \apj, 399, L59
\bibitem[Carlstrom(1990)]{carl90}Carlstrom, J. E. 1990, in ASP Conf. Ser. 12, The Evolution of the Interstellar Medium, ed. L. Blitz, 339
\bibitem[Das, Anantharamaiah \& Yun(2001)]{day01}Das, M., Anantharamaiah, K. R., \& Yun, M. S. 2001, \apj, 549, 896
\bibitem[Dyson, William \& Redman(1995)]{dwr95}Dyson, J. E., Williams, R. J. R., \& Redman, M. P. 1995, \mnras, 277, 700
\bibitem[Engelbracht et al.(1998)]{engel98}Engelbracht, C. W., Rieke, M. J., Rieke, G. H., Kelly, D. M., \& Achtermann, J. M. 1998, \apj, 505, 639
\bibitem[Forbes et al.(2000)]{forbes00}Forbes, D. A., Polehampton, E., Stevens, I. R., Brodie, J. P., \& Ward, M. J. 2 000, \mnras, 312, 689
\bibitem[Forbes et al.(1991)]{forbes91}Forbes, D. A., Ward, M. J., Depoy, D. L. 1991, \apj, 380, L63
\bibitem[Forbes et al.(1993)]{forbes93}Forbes, D. A., Ward, M. J., Rotaciuc, V., Blietz, M., Genzel, R., Drapatz, S., van der Werf, P. P., Krabbe, A. 1993, \apj, 406, L11
\bibitem[Kalas \& Wynn-Williams(1994)]{kww94}Kalas, P., \& Wynn-Williams, C. G. 1994, \apj, 434, 546
\bibitem[Keto et al.(1993)]{keto93}Keto, E., Ball, R., Arens, J., Jernigan, G., Meixner, M., Skinner, C., \& Graham, J. 1993, \apj, 413, L23
\bibitem[Keto et al.(1999)]{keto99}Keto, E., Hora, J. L., Fazio, G. G., Hoffmann, W., \& Deutsch, L. 1999, \apj, 518, 183
\bibitem[Kotilainen et al.(1996)]{koti96}Kotilainen, J. K., Forbes, D. A., Moorwood, A. F. M., van der Werf, P. P., \& Ward, M. J. 1996, \aap, 313, 771
\bibitem[Lizano et al.(1996)]{lizano96}Lizano, S., Canto, J., Garay, G., \& Hollenbach, D. 1996, \apj, 468, 739
\bibitem[Mebold et al.(1980)]{mebold80}Mebold, U., Shaver, P. A., Bell, M. B., \& Seaquist, E. R. 1980, \aap, 82, 272
\bibitem[Meurer et al.(1995)]{meurer95}Meurer, G. R., Heckman, T. M., Leitherer, C., Kinney, A., Robert, C., \& Garnett, D. R. 1995, \aj, 110, 2665
\bibitem[Mohan et al.(2001)]{mohan01}Mohan, N. R., Anantharamaiah, K. R., \& Goss, W. M. 2001, \apj, 557, 659
\bibitem[Mohan et al.(2002)]{mohan02}Mohan, N. R., Anantharamaiah, K. R., \& Goss, W. M. 2002, \apj, 574, 701
\bibitem[Paglione, Tosaki, Jackson(1995)]{paglione95}Paglione, T. A. D., Tosaki, T., \& Jackson, J. M. 1995, \apj, 454, L117
\bibitem[Payne, Anantharamaiah \& Erickson(1994)]{pae94}Payne, H. E., Anantharamaiah, K. R., \& Erickson, W. C. 1994, \apj, 430, 690 
\bibitem[Peng et al.(1996)]{peng96}Peng, R., Zhou, S., Whiteoak, J. B., Lo, K. Y., \& Sutton, E. C. 1996, \apj, 470, 821
\bibitem[Pi\~na et al.(1992)]{pina92}Pi\~na, R. K., Jones, B., Puetter, R. C., \& Stein, W. A. 1992, \apj, 401, L75
\bibitem[Puxley et al.(1997)]{puxley97}Puxley, P. J., Mountain, C. M., Brand, P. W. J. L., Moore, T. J. T., Nakai, N. 1997, \apj, 485, 143
\bibitem[Salem \& Brocklehurst(1979)]{salembro79}Salem, M., \& Brocklehurst, M. 1979, \apjs, 39, 633
\bibitem[Sams et al.(1994)]{sams94}Sams, B. J. III, Genzel, R., Eckart, A., Tacconi-Garman, L., \& Hofmann, R. 1994, \apj, 430, L33
\bibitem[Seaquist \& Bell(1977)]{sb77}Seaquist, E. R. \& Bell, M. B. 1977, \aap, 60, L1
\bibitem[Shaver, Churchwell \& Walmsley(1978)]{shaver78}Shaver, P. A., Churchwell, E., \& Walmsley, C. M. 1978, \aap, 64, 1
\bibitem[Taylor et al.(1999)]{bible}Taylor, G. B., Carilli, C. L., \& Perley, R. A. 1999, ``Synthesis Imaging in Radio Astronomy II", Sixth NRAO/NMIMT Synthesis Imaging Summer School, ASP Conference Series, Vol. 180
\bibitem[Telesco \& Harper(1980)]{th80}Telesco, C. M., \& Harper, D. A. 1980, \apj, 235, 392
\bibitem[Turner \& Ho(1985)]{th85}Turner, J. L., \& Ho, P. T. P. 1985, \apj, 299, L77
\bibitem[Ulvestad(2000)]{ulvi00}Ulvestad, J. S. 2000, \aj, 120, 278
\bibitem[Ulvestad \& Antonucci(1997)]{ua97}Ulvestad, J. S., \& Antonucci, R. R. J. 1997, \apj, 488, 621
\bibitem[Walmsley \& Watson(1982)]{ww82}Walmsley, C. M., \& Watson, W. D. 1982, \apj, 260, 317
\bibitem[Watson et al.(1996)]{watson96}Watson, A. M., et al. 1996, \aj, 112, 534
\bibitem[Weaver et al.(2002)]{weaver02}Weaver, K. A., Heckman, T. M., Strickland, D. K., \& Dahlem, M. 2002, \apj, 576, L19
\bibitem[Zhao et al.(2001)]{zhao01}Zhao, J. -H., Goss, W. M., Ulvestad, J. S., \& Anantharamaiah, K. R. 2001, in ASP Conf. Ser. 240, Gas and Galaxy Evolution, ed. J. E. Hibbard, M. Rupen, \& J. H. v. Gorkom (San Francisco: ASP), 404
\end{thebibliography}
\end{document}